\pdfoutput=1
\documentclass{emulateapj}
\usepackage{amsmath}
\usepackage{natbib}
\shortauthors{Cameron et al.}
\begin{document}

\title{Active and passive galaxies at $z\sim2$: Rest-frame optical morphologies with WFC3}

\author{E.\ Cameron\altaffilmark{1}, C.\ M.\
  Carollo\altaffilmark{1}, P.\ A.\ Oesch\altaffilmark{1}, R.\ J.\ Bouwens\altaffilmark{2,3}, G.\ D.\
  Illingworth\altaffilmark{2}, M.\ Trenti\altaffilmark{4}, I.\
  Labb\'e\altaffilmark{5}, \\ D.\ Magee\altaffilmark{2}}
\altaffiltext{1}{Department of Physics, Swiss Federal Institute of
  Technology (ETH Zurich), CH-8093 Zurich, Switzerland; cameron@phys.ethz.ch}
\altaffiltext{2}{UCO/Lick Observatory, University of California, Santa
Cruz, CA 95064, United States}
\altaffiltext{3}{Leiden Observatory, Leiden University, NL-2300 RA
  Leiden, Netherlands}
\altaffiltext{4}{University of Colorado, Center for Astrophysics and
  Space Astronomy, 389-UCB, Boulder, CO 80309, United States}
\altaffiltext{5}{Carnegie Observatories, Pasadena, CA 91101, United States}

\begin{abstract}
We  use the high angular resolution in the near-infrared of the WFC3 on HST to determine \textit{YHVz} color-color selection criteria to identify and characterize $1.5<z<3.5$ galaxies in the HUDF09 and ERS (GOODS-South) fields. The WFC3 NIR images reveal galaxies at these redshifts that were undetected in the rest-frame UV HUDF/GOODS images, as well as true centers and regular disks in galaxies classified as highly irregular in rest-frame UV light. Across the $1.5<z<2.15$ redshift range, regular disks are unveiled in the WFC3 images of $\sim$25\% of both intermediate and high mass galaxies, i.e., above $10^{10}$ $M_\odot$. Meanwhile, galaxies maintaining diffuse and/or irregular morphologies in the rest-frame optical light---i.e., not yet dynamically settled---at these epochs are almost entirely restricted to masses below $10^{11}$$M_\odot$.  In contrast at $2.25 < z < 3.5$ these diffuse and/or irregular structures overwhelmingly dominate the morphological mix in both the intermediate and high mass regimes, while no regular disks, and only a small fraction ($\sim$25\%) of smooth spheroids, are evident above $10^{11}$$M_\odot$.  Strikingly, by $1.5 < z < 2.25$ roughly 2 out of every 3 galaxies at the highest masses are spheroids.  In our small sample, the fraction of star-forming galaxies at these mass scales decreases concurrently from $\sim$60\% to $\sim$5\%.  If confirmed, this indicates that $z\sim2$ is the epoch of both the morphological transformation and quenching of star-formation which assemble the first substantial population of massive ellipticals.\end{abstract}


\section{Introduction}
The redshift interval $1.5 < z < 3.5$ represents an important
epoch for the study of galaxy formation and evolution, 
encompassing both the peak of cosmic star formation (Lilly et al.\
1996; Madau et al.\ 1996, 1998; Hopkins \& Beacom 2006; Bouwens et al.\
2007; Tresse et al.\ 2007) and the peak of active galactic nuclei
activity (Warren, Hewett, \& Osmer 1994; Fan et al.\ 2001; Croom et
al.\ 2004; Barger et al.\ 2005).  One efficient method for identifying large
samples of galaxies at these redshifts is via application of color-based selection
criteria to optical-to-near-infrared (NIR) imaging surveys.  Established selection
criteria now exist for the identification of $z \sim 3$ systems via the
Lyman break in observed UV-to-optical colors (cf.\ LBGs: Steidel et
al.\ 1996, 2003), and $1 \la z \la 3.5$ systems via the Balmer break or 4000$\AA$ break in observed optical-to-NIR colors (cf.\ EROs: Thompson et al.\ 1999; Franx et al.\ 2003; McCarthy et al.\ 2004; Van Dokkum et al.\ 2006; BX/BMs: Steidel et al.\ 2004; Adelberger et al.\ 2004; and \textit{BzK}s: Daddi et al.\ 2004, 2005).  Importantly, with optical-to-NIR color-selection criteria it is possible to sample uniformly both actively star-forming and passively-evolving systems at high redshift (e.g.\ Daddi et al.\ 2004).

The new Wide Field Camera 3 (WFC3) on the Hubble Space Telescope (HST) has greatly enhanced the UV and NIR imaging capabilities of this
observatory, offering sharper imaging and greater throughput with a
new complement of narrow- and broad-band filters.  In particular, the unprecedented depth and sharpness 
of the NIR WFC3 images further enhances our capacity to detect LBGs at the era of peak star formation in the universe---as well as at much higher redshifts; see, e.g.,  the WFC3 detections of
numerous galaxy candidates at $z \sim 7$-8
(Bouwens et al.\ 2010a,b; Oesch etal.\ 2010; Wilkins et al.\ 2010; McLure et al.\ 2010; Bunker et al.\ 2010; Finkelstein et al.\ 2010), and even $z \sim10$ (Bouwens et al.\ 2011).

 In this paper we use the WFC3 data acquired so far in the context of the Hubble Ultra Deep Field 2009 (HUDF09) and
Early Release Science (ERS) programs to \textsc{(i)}  extract $H_{160}$-band based galaxy catalogs down to, respectively, AB=27 and 25 mag;  \textsc{(ii)} derive  optimal
color-selection criteria (\textit{YHVz}) for the identification
and characterization of galaxies at
$1.5 < z < 3.5$,  based on the WFC3 NIR filter set (complemented by optical ACS data),  and \textsc{(iii)} use the constructed WFC3 $H_{160}$-band source catalogs  to investigate, through qualitative comparisons with archival ACS optical data in the HUDF09 and ERS fields, differences and similarities in the rest-frame UV and optical morphologies  of $1.5 <z<3.5$ galaxies.
The efficiencies of the \textit{YHVz} selection criteria presented here are established via comparisons with the spectroscopic and photometric redshifts
of galaxies in both datasets.  Alternative color-selection criteria, \textit{VIH} and \textit{JHVI}, relevant to the  filter set of CANDELS/Wide, are also explored using the ERS dataset.  

The outline of the paper is as follows.  In Section \ref{data} we describe
the imaging data now available from the  HUDF09 and ERS
observational programs, and briefly review
the other datasets from which additional UV-to-IR 
magnitudes and spectroscopic redshifts are obtained.  In Section \ref{method}
we detail the construction of our master source catalogs based on the
HUDF09 and ERS $H_{160}$ imaging, and the matching of these catalogs to multi-wavelength
photometry from previous studies.  In Section \ref{sec4} we describe the
estimation of photometric redshifts, specific star formation rates, and stellar masses via spectral energy distribution
(SED) fitting.  In Section \ref{results} we present \textit{YHVz}
color-selection criteria for the identification of galaxies at $1.5
< z < 3.5$ using the WFC3 and ACS filters.  In addition, we carefully quantify selection efficiencies and contamination rates for these criteria according to the spectroscopic, and
(up to) 15 band photometric, redshifts of
galaxies in our master source catalogs.  Furthermore, the characterization of
passively-evolving galaxies at $1.5 < z < 3.5$ via a simple
extension of our selection criteria (\textit{passive YHVz}) is
explored in Section \ref{pYHVz}, and the alternative color-selections \textit{JHVI} and \textit{VIH}, using filters from the CANDELS/Wide program (excluding $Y_{105}$ which is available only within the GOODS-N field; Koekemoer et al.\ 2011), are presented in Section \ref{mct}.  In Section \ref{sourcecatalogs} we publish our master source catalogs containing positions, photometric apertures, and photometric redshifts for all robust detections to $H_{160} < 27$ mag in the HUDF09 and $H_{160} < 25$ mag in the ERS.  In Section \ref{morph} we compare the rest-frame UV and optical morphologies of galaxies in the HUDF09 and ERS fields at $1.5 < z  <3.5$, and examine the mass scales over which key structural sub-types are revealed by the WFC3.  We summarize in Section
\ref{Conclusions}.  Note that we adopt a
cosmological model with $\Omega_\Lambda = 0.75$, $\Omega_M = 0.25$,
and $h=0.7$, and that all magnitudes are quoted in the AB system throughout.

\section{Data}\label{data}
\subsection{HUDF09}
The HUDF09 program (Bouwens et al.\ 2010) now in progress will observe
a $\sim$4.7 arcmin$^2$ central region of the HUDF (Beckwith et al.\
2006) with the new Wide Field Camera 3 (WFC3) on the
\textit{Hubble Space Telescope} (HST)
for a total of 96 orbits\footnote{A further 96 orbits in the HUDF09
  program will be dedicated to imaging, at similar depth, of the
two nearby HUDF05 fields (Oesch et al.\ 2007, 2009).}.  The exposure time
is divided between three NIR filters: $Y_{105}$ (F105W),
$J_{125}$ (F125W), and $H_{160}$ (F160W).  The first epoch data
used here total 60 orbits (16 in $Y_{105}$, 16 in $J_{125}$,
and 28 in $H_{160}$), and reach a uniform depth in all filters of
$\sim$28.8 mag in a $0.2$ arcsec radius aperture (despite the necessary rejection of two orbits of $Y_{105}$
imaging due to severe persistence issues).  Our reduction of the
WFC3 imaging data using the
\texttt{multidrizzle} package (Koekemoer et al.\ 2002) produces final science grade images with PSF
FWHM of $\sim$0.15 arcsec and a pixel size of 0.06 arcsec (see Bouwens et al.\ 2010).   Rebinning of the HUDF ACS $B_{435}$ (F435W), $V_{606}$ (F606W), $i_{775}$
(F775W), and $z_{850}$ (F850LP) 
images to the scale of our WFC3 frames was also performed to facilitate
aperture-matched photometry.

\subsection{ERS}
The Early Release Science (ERS) program 
has observed a northern section of the GOODS-South field (Giavalisco
et al.\ 2004) in the WFC3/IR channel in 10 pointings for a total of 60
orbits (cf.\ Windhorst et al.\ 2010).  The exposure time was divided evenly between three NIR
filters: $Y_{098}$ (F098M), $J_{125}$ (F125W), and $H_{160}$ (F160W),
reaching a depth of $\sim$27.5 mag (in a 0.2 arcsec radius aperture)
over an area of $\sim$40 arcmin$^2$.  Our reduction of this data (Bouwens et al.\ 2010) was
performed using the same procedures as for the HUDF09 WFC3 imaging,
resulting in final science grade images with PSF FWHM $\sim$0.15
arcsec and a pixel size of 0.06 arcsec.  The GOODS ACS $B_{435}$ (F435W), $V_{606}$ (F606W), $i_{775}$
(F775W), and $z_{850}$ (F850LP) images were again rebinned to the same scale as the reduced WFC3 frames.

\subsection{Ancillary Datasets}
A wealth of complementary data, including photometry at wavelengths beyond the range of
the HUDF/HUDF09 and GOODS/ERS HST imaging has been gathered for objects in the HUDF and GOODS-South
fields by a number of different teams.  Much of this data
where publicly available has been compiled in the GOODS-MUSIC
catalog (Grazian et al.\ 2006; Santini et al.\ 2009), which we employ in this study for
objects with unambiguous matches in our primary source catalogs (see
Section \ref{catalog} below).  The GOODS-MUSIC catalog
contains PSF- and aperture-matched photometry (derived via the
\texttt{ConvPhot} package; De Santis et al.\ 2007) in the following
additional filters: ESO 2.2-WFI and VLT-VIMOS $U$, ESO VLT/ISAAC $J$, $H$,
and $K_\mathrm{s}$, Spitzer IRAC 3.6, 4.5, 5.8, and 8.0 $\mu$m, and MIPS  24 $\mu$m.  A compilation of spectroscopic
redshifts is also included in the GOODS-MUSIC catalog, mainly sourced from the GOODS
(Vanzella et al.\ 2005, 2006, 2008), K20 (Mignoli et al.\ 2005), and VVDS
(Le F\`evre et al.\ 2004) projects.

In addition to the GOODS-MUSIC catalog, we also employ the published
spectroscopic redshifts and spectral classifications for passive galaxies at $z \sim 2$ in the HUDF
and GOODS-South fields from Daddi et al.\ (2005) and
Cimatti et al.\ (2008) (hereafter \textit{pBzK}).  The former sample consists of 7 high redshift sources in the HUDF
with \textit{passive  BzK} colors, confirmed via analysis of low resolution GRAPES spectra, of which 3 lie in the region of WFC3/IR imaging from the HUDF09
program.  The latter  sample  consists of 13 galaxies
from the GMASS catalog of Spitzer 4.5 $\mu$m detected
sources in the GOODS-South field with a prominent Mg$_\mathrm{UV}$
feature in their ESO VLT/FORS2 spectra, of which 10 lie in the
region of WFC3/IR imaging from the ERS program.  In total, 2 and 8 of
these \textit{pBzK} galaxies in the Daddi et al.\ and Cimatti et al.\ samples, respectively,
lie within the target range of our \textit{YHVz}
color-selection criteria ($1.5 < z < 3.5$).
A further 3 and 6 GMOS spectroscopic redshifts from the $1 < z < 2$
Roche et al.\ (2006) ERO sample were employed for objects in the
HUDF09 and ERS, respectively.  Although only one of the corresponding
objects was missing a `confident' spectroscopic redshift in the
GOODS-MUSIC catalog, the Roche et al.\ redshifts were preferred due to the higher
confidence and precision of these measurements.  Importantly, for one object (ERS0325) 
assigned $z_\mathrm{spec}=0.481$ in the GOODS-MUSIC catalog, the value
of $z_\mathrm{spec}=2.017$ reported by the Roche et al.\ team is much
more consistent with its observed UV-to-IR SED.  Finally, for the purposes of star-galaxy separation we
also refer to the PEARS-S star catalog of Pirzkal et al.\ (2009).

\section{Extraction of the $H_{160}$ HUDF09 and ERS Source Catalogs}\label{method}

\subsection{Details}\label{catalog}
Master source catalogs based on all robust $H_{160}$ detections to $H_{160} <
27$ mag and $H_{160} < 25$ mag in the HUDF09 and ERS imaging, respectively, were
constructed as follows.  First, \texttt{SExtractor} (Bertin \& Arnouts
1996) was run on each of the reduced $H_{160}$ science images with four
different choices of deblend parameters representing \textit{hot}
(\texttt{nthresh} $=$ 64), \textit{medium} (\texttt{nthresh} $=$ 16),
\textit{cold} (\texttt{nthresh} $=$ 8), and \textit{very cold} (\texttt{nthresh} $=$ 2) extractions.   In each
case, the RMS frames output from \texttt{multidrizzle} were scaled to
reflect the true signal-to-noise of the reduced science images before their application as weight maps, and the following key control parameters
were adopted: \texttt{detect\_thresh} $=$ $1.5\sigma$,
\texttt{detect\_minarea} $=$ 12, and \texttt{min\_cont} $=$
0.0001.  Based on these $H_{160}$
extractions we then re-ran \texttt{SExtractor} in dual image mode to recover
aperture-matched fluxes and Kron magnitudes from the remaining optical-to-NIR
HST images ($B_{435}$ to $J_{125}$).  Careful visual inspection of the output Kron ellipses of 
sources identified in each extraction was then performed to remove
spurious detections, and to identify
the most appropriate aperture for flux measurement of each real
source.  When choosing the best aperture we considered both the
positions of possible matches in the
GOODS-MUSIC catalog and the appearance of each source in all seven ACS
plus WFC3 ($B_{435}$ through $H_{160}$) filters.  In ambiguous cases
we favoured the
deblending of objects into distinct concentrations of $H_{160}$ flux (the
filter expected to best trace the underlying stellar mass
distribution, modulo some uncertainty regarding the role of TP-AGB stars).  For calibration of the NIR
magnitudes we
employed the latest (AB system) WFC3/IR zero-points available from the STSCi
webpage\footnote{\texttt{URL(http://www.stsci.edu/hst/wfc3/phot\_zp\_lbn)}}, and
for the optical magnitudes we adopted the ACS zero-points determined
by the GOODS team\footnote{\texttt{URL(http://archive.stsci.edu/pub/hlsp/goods/v2/)}}.
  A total of 1052 and 3078 sources were thereby recovered to $H_{160}
  < 27$ mag
  and $H_{160} < 25$ mag in the HUDF09 and ERS imaging, respectively.  To
  define the
  master source catalogs employed in this analysis we limit our
  samples to those systems with high
S/N coverage in all 7 available ACS plus WFC3
filters\footnote{Identified as those objects occupying areas of the reduced
  science images with local background RMS less than twice the global
  background RMS.} and lying within
the region of sky covered by the GOODS-MUSIC catalog, a total of 993
and 2630 objects in the HUDF09 and ERS, respectively.

For all objects in our  master source catalogs we searched for counterparts
in the GOODS-MUSIC catalog (Santini et al.\ 2009) with matching
positions and photometric apertures.  All object pairs with centroid differences of $<$0.24 arcsec (i.e., 4
pixels in our drizzled WFC3 images) and $B_{435}$, $V_{606}$, $i_{775}$, and $z_{850}$ magnitudes
differing by less than 0.5 mag between the two catalogs were
classified automatically as successfully matched---426 sources in the HUDF09
and 1954 in the ERS fields.  The remaining object pairs not satisfying these
criteria were visually inspected, and a further 44 and 294
matches were thereby recovered for the HUDF09 and ERS fields, respectively.

For most galaxies at $H_{160}>24.5$ mag no counterpart could be
identified due to the brighter
GOODS-MUSIC selection limits ($z_{850} < 26$ mag or $K_\mathrm{s} < 23.5$
mag).  In total, 523 and 382 objects in our master source catalogs from the
HUDF09 and ERS, respectively, were ultimately deemed to be missing
valid counterparts in the GOODS-MUSIC catalog.

\begin{figure}
\epsscale{1.15}
\plotone{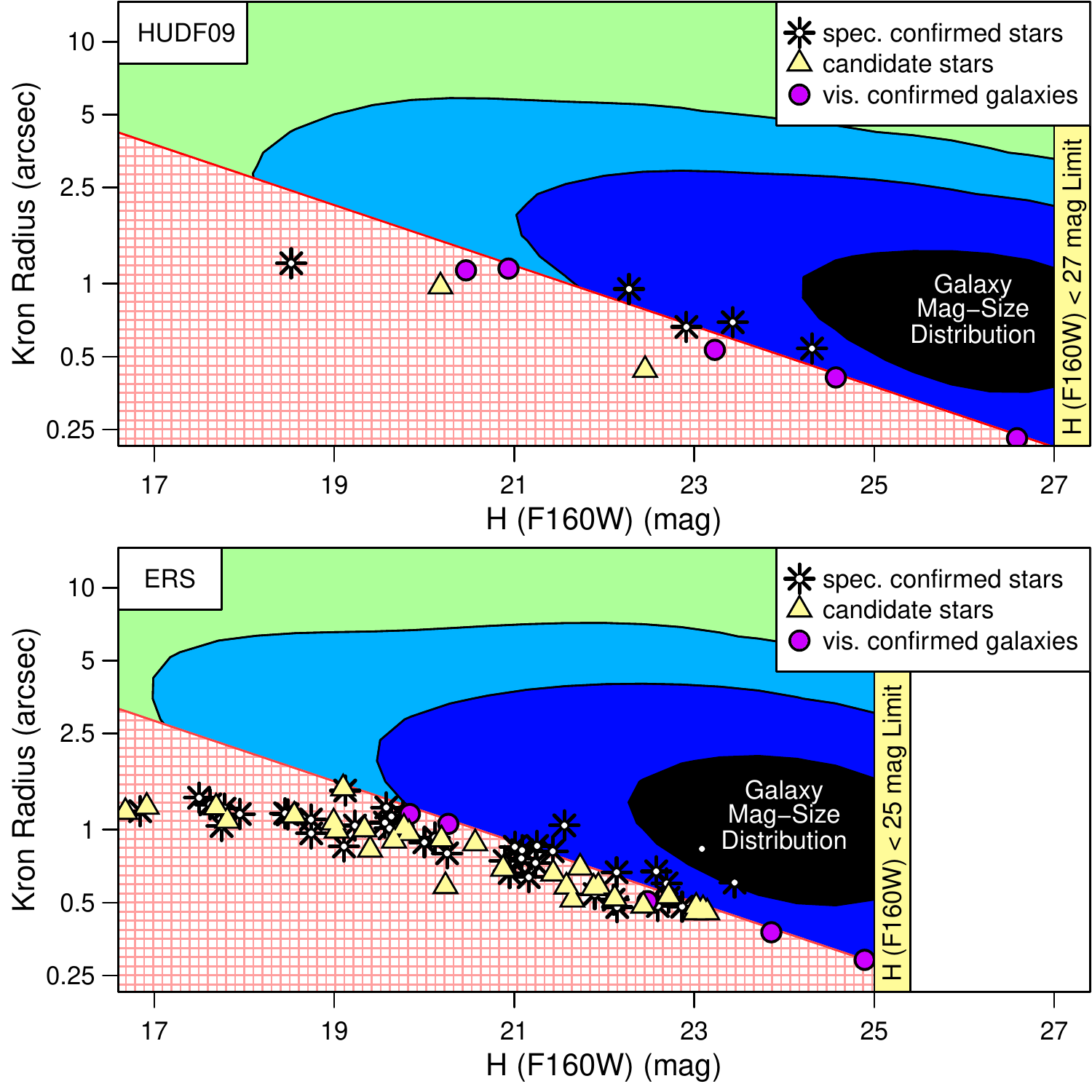}
\caption{The distributions of galaxies, spectroscopically-confirmed stars, and candidate stars in the $H_{160}$ apparent magnitude$-$Kron size plane in our master source catalogs to $H_{160} < 27$ mag in the HUDF09 (\textit{top row}) and $H_{160} < 25$ mag in the ERS (\textit{bottom row}).  The red hashed area in each case denotes the region within which small size outliers were classified as candidate stars unless resolved structure was clearly evident upon visual inspection (as described in Section \ref{sg}).  The black, dark blue, and light blue (smoothed) contours respectively enclose 68\%, 95\%, and 99\% of the detected objects in each survey.}
\label{fig1}
\end{figure}

\subsection{Star-Galaxy Separation}
\label{sg}

\begin{figure*}
\epsscale{1.15}
\plotone{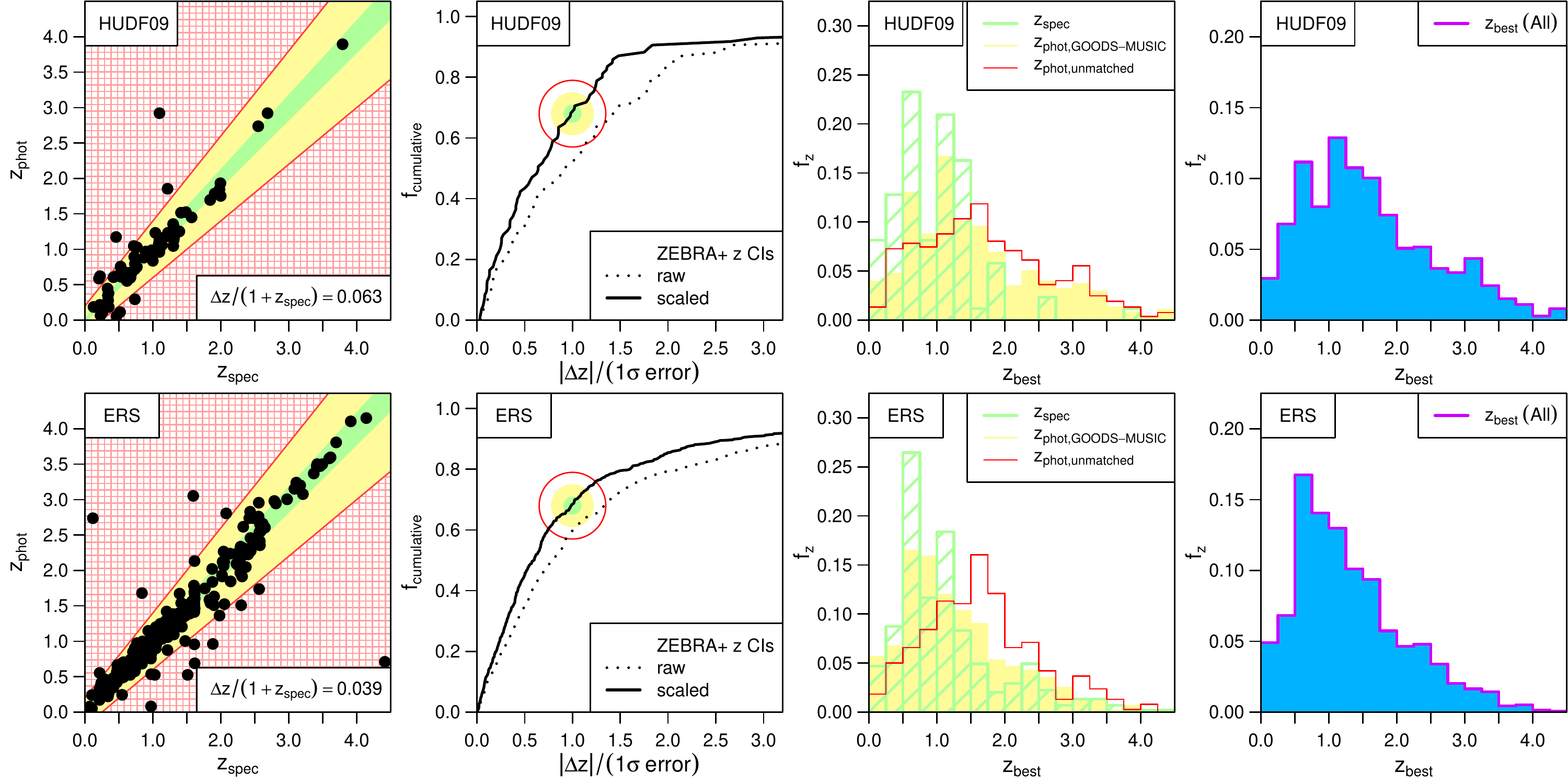}
\caption{Evaluation of the accuracy of our ZEBRA photometric redshifts for galaxies in the HUDF09 (\textit{top row}) and ERS (\textit{bottom row}).  The panels in the left most columns present the one-to-one comparison between our photometric redshifts and available spectroscopic redshifts, while those in the colums second from the left evaluate our estimates of the photometric redshift uncertainties for these galaxies against the same spectroscopic benchmarks.  (The green, yellow, and red shaded regions here mark deviations of $<5$\%, $<20$\%, and $>20$\% in $\Delta z / (1+z)$, respectively.)  The panels in the two right most columns illustrate the redshift distributions of galaxies in each field both as an ensemble, and after subdivision into those systems with spectroscopic redshifts, photometric redshifts with matched GOODS-MUSIC photometry, and HST-based photometric redshifts only.  See Section \ref{photzs} for further details.}
\label{fig2}
\end{figure*}

Star-galaxy separation in our HUDF09 and ERS samples was performed as follows.  First, all sources matched to counterparts in the
GOODS-MUSIC catalog with spectroscopic redshift quality flag $<$2 and $z_\mathrm{spec}=0$ (2
objects in the HUDF09, and 29 in the ERS) and all objects with centroids matched
within $0.36$ arcsec to members of the PEARS-S star catalog of
Pirzkal et al.\ (2009) (an additional 3 and 16 sources in the HUDF09
and ERS, respectively) were identified as spectroscopically-confirmed
stars.  In addition, all ultra-compact outliers from the observed $H_{160}$
apparent magnitude$-$Kron radius distributions at $\log_{10}
(R_\mathrm{Kron}/'') < -0.125 \times (H_{160}-20)+\gamma$ (with $\gamma =$ 0.2 for the HUDF09 and 0.075 for the ERS) not already
identified as spectroscopically-confirmed stars were
visually-inspected and classified as candidate
stars (2 in the HUDF09, and 35 in the ERS) unless resolved structure was clearly evident.  Figure \ref{fig1} shows these candidate stars on the $H_{160}$
apparent magnitude$-$Kron radius diagnostic diagram.  We exclude these sources from
our analysis of galaxy color-selection efficiencies in Section \ref{results}.

\section{Estimates for redshifts, specific star formation rates, and stellar masses}
\label{sec4}
\subsection{Photometric Redshift Estimates}
\label{photzs}
We have computed photometric redshifts for all objects in our HUDF09 and ERS source catalogs via
template fitting to the observed spectral energy distributions in the
HST ACS and WFC3 filters ($B_{435}$ to $H_{160}$), supplemented by the
additional photometry in the GOODS-MUSIC catalog (for a total of 15 passbands) where a valid counterpart
could be identified.  The actual template fitting was
performed using the \textit{Zurich
  Extragalactic Bayesian Redshift Analyzer} (ZEBRA; Feldmann et al.\
2006).  To recover optimal photometric
redshifts ZEBRA modifies an empirical template library (Polletta et
al.\ 2007; plus model SEDs for very blue objects from Bruzual \& Charlot 2003) to account for systematic mismatches
against the observed spectral energy distributions. 

We quantify empirically our mean photometric redshift uncertainties by comparison against matched objects with confident (quality flag $<$2) spectroscopic redshifts in the GOODS-MUSIC, Daddi et al., Cimatti et al., and Roche et al.\ catalogs (86 in the
HUDF09, and 446 in the ERS), as illustrated in the panels in the left most columns of Fig.\ \ref{fig2}.  We thereby estimate typical uncertainties ($\Delta z = |z_\mathrm{spec}-z_\mathrm{phot}|$) of $\Delta z \sim 0.063(1+z)$ and $\Delta z \sim 0.039(1+z)$ for the HUDF09 and ERS master source catalogs, respectively.  The larger uncertainties in the HUDF09 photometric redshifts reflect the relatively small number of confident spectroscopic measurements available in this field for calibration of our SED template library.  We also note low rates of catastrophic failures for the ZEBRA photometric redshift pipeline, namely that $<$5\% of galaxies in the HUDF09, and $<$9\% of galaxies in the ERS, are in disagreement with their spectroscopic redshifts by $\Delta z > 0.2(1+z)$.

At $z < 1.5$, where the greatest numbers of spectroscopic redshifts are available (78 in the HUDF09, and 355 in the ERS), the mean uncertainties are reduced to $\Delta z \sim 0.058(1+z)$, and $\Delta z \sim 0.035(1+z)$, respectively.  Nevertheless, at $z > 1.5$ our photometric redshift accuracy remains satisfactory with uncertainties of only $\Delta z \sim 0.069(1+z)$, and $\Delta z \sim 0.056(1+z)$, respectively---although we note that in the case of the HUDF09 the reliability of this comparison is limited by the small number (8) of spectroscopic redshifts in this interval.  Hence, as a further stage of quality control on our ZEBRA photometric redshifts we also evaluate here the performance of the output confidence intervals derived from the posterior probability distribution in redshift space for each galaxy.  Specifically, we examine the ratio between $\Delta z$ (i.e., the absolute difference between the maximum likelihood photometric redshift and the true spectroscopic redshift) and the width of the one sigma confidence interval for each galaxy (cf.\ Ilbert et al.\ 2009).  If our posterior probability distributions are well estimated then we would expect the value of this quotient to be less than one for roughly 68\% of each sample.  In fact, our investigation reveals a slight underestimation of uncertainties by ZEBRA such that a rescaling of these confidence interval widths by 1.25 and 1.35 for the HUDF09 and ERS samples, respectively, is required in order to achieve the 68\% target.  Having calibrated our photometric redshift uncertainties we fold these via Monte Carlo simulation into our subsequent error calculations for the selection efficiencies of the color-color criteria presented in Section \ref{results}.

In Fig.\ \ref{fig2} we summarize graphically the key properties of our HUDF09 and ERS galaxy redshift catalogs.   In order from the lefthand side of the page the four columns of this figure each contain two panels demonstrating for the HUDF09  and ERS (top and bottom rows, respectively): \textsc{(i)} the close agreement between our photometric redshift estimates and the benchmark spectroscopic values where available in each sample; \textsc{(ii)} the necessity of our (slight) re-scaling of the (ZEBRA) estimated uncertainties to ensure consistency in the cumulative distribution functions for (absolute) empirical redshift error divided by one sigma confidence interval width, i.e., $|\Delta z|$/(1$\sigma$ error); \textsc{(iii)} relative frequency histograms contrasting the redshift distributions of galaxies with spectroscopic redshifts against those for galaxies without (further subdivided according to presence/absence of additional GOODS-MUSIC photometry)---the latter distinction thereby highlighting the unique power of WFC3/IR imaging for exploring the $z \sim 1.5$ discovery space; and \textsc{(iv)} the overall redshift distributions for all galaxies in our magnitude limited samples---both of which peak around $z \sim 1$, as is typical for untargeted, deep imaging surveys with optical/near-IR selection (e.g.\ Rudnick et al.\ 2001; Ilbert et al.\ 2009).

Finally, we note that our HUDF09 $H_{160} < 27$ mag master source catalog contains both $z \sim 7$ candidates at this brightness from Oesch et al.\ (2010) (UDFz-42566566 and UDFz-38807073) plus the probable supernova (UDFz-34537360), all of which were assigned photometric
redshifts $z\sim6.9$ by ZEBRA.  However, the remainder of the recently reported $z \sim 7$ and $z \sim 8$ candidates in the HUDF09 and ERS fields (Bouwens et al.\ 2010a,b; Oesch et al.\ 2010; Wilkins et al.\ 2010) are fainter than the apparent magnitude limits employed in this study.

\subsection{Specific Star Formation Rate and Stellar Mass Estimates}\label{ssfrs}
Specific star formation rates (SSFRs) and stellar masses for all galaxies in our HUDF09 and ERS
samples were estimated using an updated version of ZEBRA, ZEBRA+ (Oesch
et al.\ 2010, in prep.).  ZEBRA+ computes various galaxy
physical properties based on template fitting to model SEDs with the
(previously-computed) photometric redshift, or measured spectroscopic
redshift, held fixed.  For this analysis we employed a library of SED templates
based on the Bruzual \&
Charlot (2003) stellar population synthesis code with a Chabrier IMF (Chabrier 2003), drawn from a grid
of exponential decay star formation rate models well-sampled over a wide range of
ages (0.01$-$12 Gyr), decay rates ($\tau$ $=$ 0.05$-$9 Gyr), and
metallicities (0.05$-$2 $Z_\odot$).  The impact of dust
reddening is accounted for during template fitting via a
Calzetti extinction law (Calzetti 2001) with the E$(B-V)$ value a free parameter (see Sargent et al.\ 2010 for a demonstration of the impact of dust reddening on galaxies at intermediate redshift).  Of course, there is a substantial degree of degeneracy in the choice of stellar population template and dust extinction model inherent in the fitting of broad-band SEDs, which is expected from previous studies to dominate the error budget (contributing $\sigma_{\log M} \approx 0.20$ dex) in samples for which broad-band flux measurements are available in a significant number of filters adequately sampling each galaxy SED (e.g.\ Bolzonella et al.\ 2010; Taylor et al.\ 2010).  Thus, for galaxies in our sample matched to counterparts in the GOODS-MUSIC survey we expect to achieve roughly this level of accuracy, although for galaxies with HST-only photometry our uncertainties will be necessarily higher.  Details for ZEBRA+, including a further investigation into the relevant uncertainties, are given in Oesch et al.\ (2011, in prep.; see also Oesch et al.\ 2009).

We note that of the 3 and 10  spectroscopically-confirmed passive galaxies of Daddi et al.\ and Cimatti et al., respectively, which enter our HUDF09 and ERS samples, all but one have ZEBRA+ estimated SSFRs representative of passively-evolving stellar populations (i.e., SSFR $< 10^{-10}$ yr$^{-1}$).  The one discrepant object (ERS2953) was assigned an intermediate SSFR of $1.2 \times 10^{-9}$ yr$^{-1}$ with an extreme E$(B-V)$ fit of 0.6 mag.

\begin{figure}
\epsscale{1.15}
\plotone{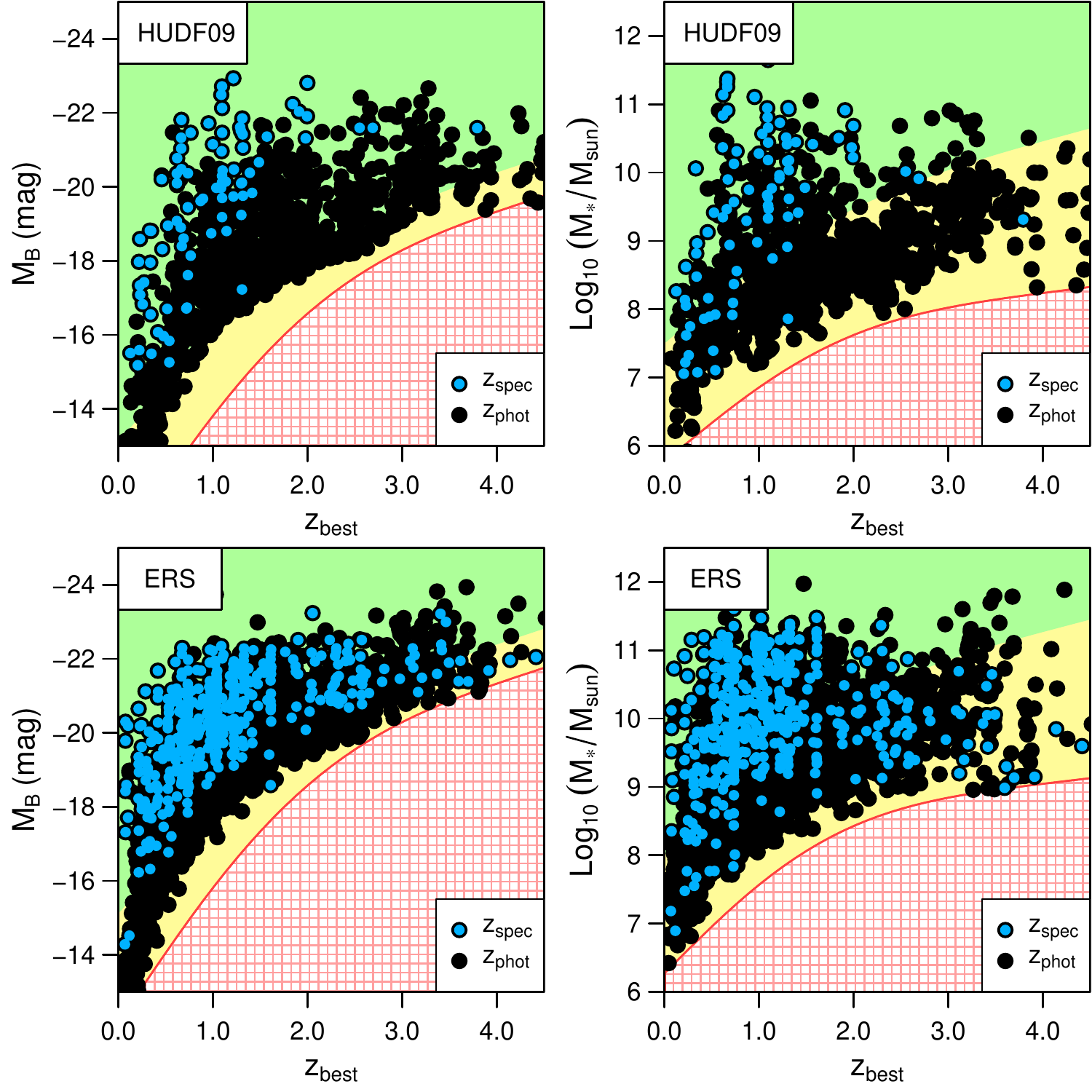}
\caption{The distributions of galaxies in absolute (rest-frame $B$) magnitude and stellar mass derived via the ZEBRA+ code for galaxies in our master source catalogs to $H_{160} < 27$ mag in the HUDF09 (\textit{top row}) and $H_{160} < 25$ mag in the ERS (\textit{bottom row}).  The (conservative) SED-\textit{independent} selection limits for each sample are indicated via the green-shaded regions on each plot, while the boundaries of SED-\textit{dependent} incompleteness (set by the difference between the reddest and bluest spectral templates matched in our fits) are marked in yellow (with red hashes indicating 100\% detection incompleteness).  Galaxies in our samples for which reliable spectroscopic redshift measurements are available are highlighted in blue.}
\label{fig3}
\end{figure}

In Fig.\ \ref{fig3} we present the distributions of galaxies in our HUDF09 and ERS master source catalogs in both absolute (rest-frame $B$-band) magnitude and stellar mass as function of redshift.  Two selection limits are indicated on each plot: \textsc{(i)} the SED-\textit{dependent} completeness limit corresponding to the lower bound on luminosity/mass of the faintest/least massive systems detectable at a given redshift above our $H_{160}$ magnitude threshold---all members of our HUDF09 and ERS samples thus lie above, or on, this line; and \textsc{(ii)} the SED-\textit{independent} completeness limit corresponding to the respective lower bound under the constraint of full detectability for galaxies of \textit{all} spectral types.  Note that the observed distributions of galaxies are not tightly bound by the former until $z \sim 2.5$ as the largest K-correction from observed $H_{160}$ to rest-frame $B$ below this redshift occurs for the reddest spectral type (which is not matched to any of the faintest/least massive systems in these surveys; see, for comparison, Cameron \& Driver 2007, 2009).  The SED-\textit{independent} limits presented here set the thresholds ($M > 10^{10}$$M_\odot$ for the HUDF09 and $M > 10^{11}$$M_\odot$ for the ERS) for selection of the mass-limited galaxy samples used to investigate the morphological mix at $z \sim 2$ in Section \ref{morph}.

\section{Optimization of HST-based color-selection criteria for $1.5<\lowercase{z}<3.5$ galaxies}\label{results}

\subsection{\textit{YHV\lowercase{z}} Color-Selection}\label{colorcolor}
\begin{figure*}
\epsscale{1.15}
\plotone{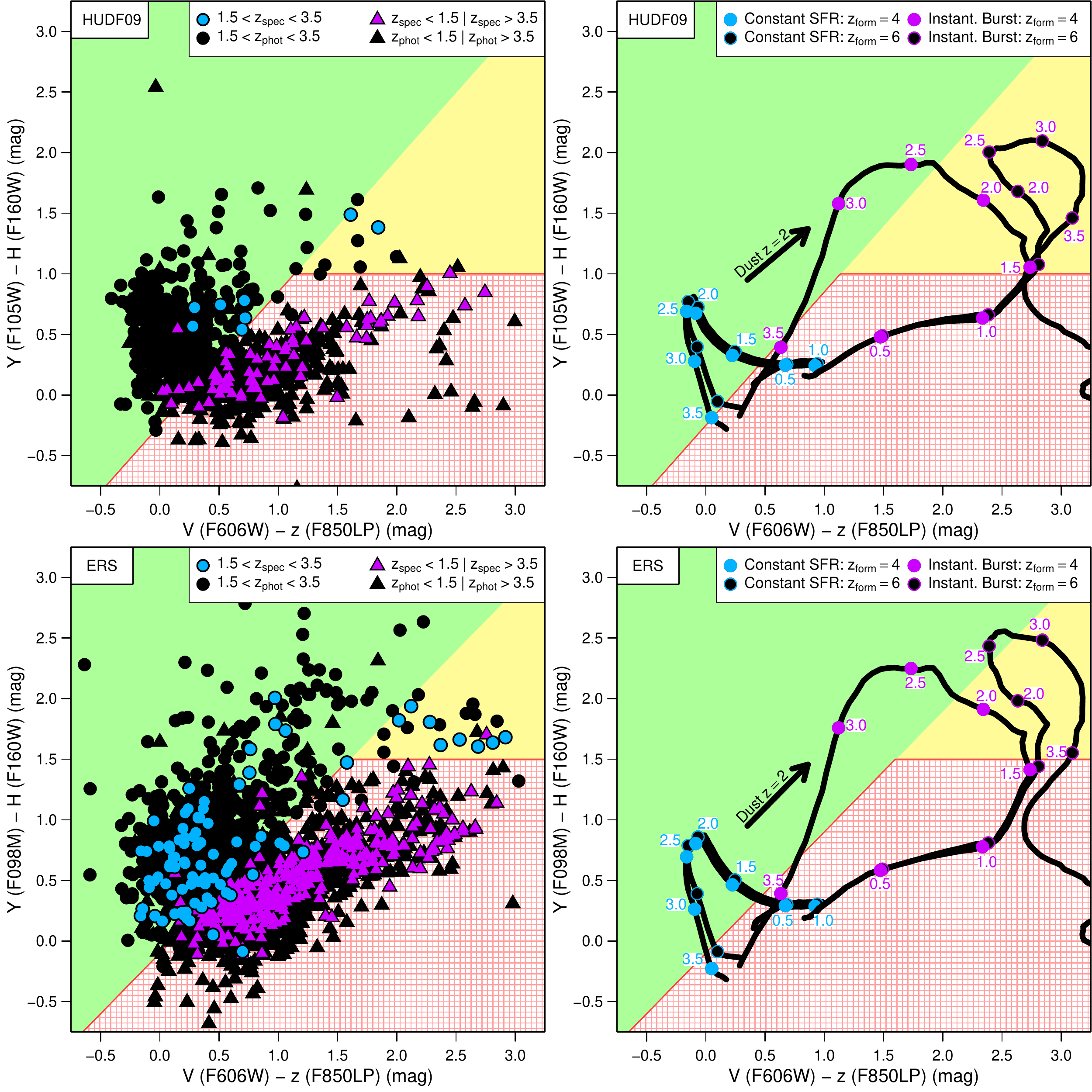}
\caption{\textit{YHVz} color-selection of galaxies at $1.5 < z < 3.5$ in the \textit{Hubble Space Telescope} ACS plus WFC3 imaging of the HUDF and GOODS-South fields from the HUDF09 (\textit{top row}) and ERS (\textit{bottom row}) programs.  The observed distributions in $Y_{105/098}-H_{160}$ vs.\ $V_{606}-z_{850}$ color-color space of all objects in our HUDF09 ($H_{160} < 27$ mag) and ERS ($H_{160} < 25$ mag) master source catalogs are shown in the left hand panels.  Galaxies with redshifts in the range $1.5 < z < 3.5$ are marked with circles, while those outside this redshift interval are marked with triangles, and these symbols are highlighted blue and purple, respectively, if the redshift is spectroscopic, rather than photometric.  The optimal \textit{YHVz} selection criteria we derive for the high redshift population in each survey are indicated by the union of the green- and yellow-shaded regions---the particular significance of the yellow-shaded corner for passive galaxy selection is discussed in Section \ref{pYHVz}.  Synthetic color-color tracks revealing the redshift evolution of archetypal active and passive stellar population models (with $Z=0.35$$Z_\odot$, and both $z_\mathrm{form}=4$ and $z_\mathrm{form}=6$) in the relevant filters are shown for comparison in the right hand panels.  Example reddening vectors corresponding to Calzetti (2001) extinction for galaxies observed at $z=2$ are marked by black arrows.}
\label{fig4}
\end{figure*}

Through visual inspection of synthetic evolutionary tracks for model stellar populations in our ZEBRA+ template library we have identified $Y_{105/098}-H_{160}$ vs.\ $V_{606}-z_{850}$ as the most favorable parameter space for selection and characterization of $1.5 < z < 3.5$ galaxies given the set of WFC3 and ACS filters employed in the HUDF09/HUDF and ERS/GOODS imaging programs.  Example tracks for instantaneous burst and constant star formation rate models (with metallicities of $Z=0.35$ $Z_\odot$ and formation redshifts of both $z_\mathrm{form}=4$ and $z_\mathrm{form}=6$) are presented in Fig.\ \ref{fig4} to illustrate the
expected evolution of active and passive systems in this plane.  These tracks indicate that galaxies at $1.5 < z < 3.5$ should be significantly redder in $Y_{105/098}-H_{160}$ 
, at fixed $V_{606}-z_{850}$, than galaxies at lower or higher redshifts.  Moreover, according to the Calzetti (2001) reddening law, the effect of dust extinction on the observed galaxy colors in this diagram is simply a diagonal shift (as indicated in Fig.\ \ref{fig4}) with the potential to move lower redshift objects into the region of high redshift color-color space only for passively-evolving systems at $z \ga 1$.  When the corresponding color-color distributions of real galaxies at $1.5 < z  < 3.5$ in the HUDF09 and ERS are contrasted against
those at lower and higher redshifts (see Fig.\ \ref{fig4} again) it is clear that galaxies at these epochs are indeed well separated, occupying distinct color-color sequences with minimal overlap.

\begin{table*}
\caption{Optimal \textit{YHV\lowercase{z}} $1.5 < z < 3.5$ Selection Criteria and Corresponding Selection and Contamination Rates as a Function of $H_{160}$ Magnitude, and Comparison Against \textit{BzK} Selection$^{\dagger}$}
\begin{center}
\label{table1}
\vskip-0.5cm
\begin{tabular}{llccccc}
\\
\hline \hline
Survey & $H_{160}$ Limit & $N_\mathrm{gal}$ &  Selection Criteria & $S$ (\%)$^{\star}$ & $C$ (\%)$^{\star}$ & $C_{z<1}$ (\%)\\
\hline
HUDF09 & $H_{160} < 25$ mag & 309 & $Y_{105} - H_{160} > 1.1 \times (V_{606}
- z_{850}) - 0.25 \mathrm{\ or\ } Y_{105} - H_{160} > 1.0$  & $94\pm2$ & $17\pm4$ & $8\pm2$\\
HUDF09 & $H_{160} < 26$ mag & 540 & $Y_{105} - H_{160} > 1.1 \times (V_{606}
- z_{850}) - 0.25 \mathrm{\ or\ } Y_{105} - H_{160} > 1.0$  & $95\pm2$ & $13\pm3$ & $6\pm2$\\
HUDF09 & $H_{160} < 27$ mag & 986 & $Y_{105} - H_{160} > 1.1 \times (V_{606}
- z_{850}) - 0.25 \mathrm{\ or\ } Y_{105} - H_{160} > 1.0$  & $96\pm1$ & $15\pm2$ & $5\pm1$\\
\hline
HUDF09 & $H_{160} < 25$ mag & 260 & $(z_{850}-K_\mathrm{s}) - (B_{435} - z_{850}) > - 0.2
\mathrm{\ or\ } z_{850} - K_\mathrm{s} > 2.5$ & $88\pm3$ & $34\pm3$ & $2\pm2$\\
HUDF09 & $H_{160} < 26$ mag & 395 & $(z_{850}-K_\mathrm{s}) - (B_{435} - z_{850}) > - 0.2
\mathrm{\ or\ } z_{850} - K_\mathrm{s} > 2.5$ & $85\pm2$ & $30\pm3$ & $3\pm1$\\
HUDF09 & $H_{160} < 27$ mag & 446 & $(z_{850}-K_\mathrm{s}) - (B_{435} - z_{850}) > - 0.2
\mathrm{\ or\ } z_{850} - K_\mathrm{s} > 2.5$ & $84\pm1$ & $30\pm2$ & $3\pm1$\\
\hline
ERS & $H_{160} < 23$ mag & 680 & $Y_{098} - H_{160} > V_{606} - z_{850} - 0.1
\mathrm{\ or\ } Y_{098} - H_{160} > 1.5$ & $88\pm4$ & $21\pm4$ & $10\pm3$\\
ERS & $H_{160} < 24$ mag & 1366 & $Y_{098} - H_{160} > V_{606} - z_{850} - 0.1
\mathrm{\ or\ } Y_{098} - H_{160} > 1.5$ & $92\pm2$ & $15\pm2$ & $5\pm1$\\
ERS & $H_{160} < 25$ mag & 2563 & $Y_{098} - H_{160} > V_{606} - z_{850} - 0.1
\mathrm{\ or\ } Y_{098} - H_{160} > 1.5$ & $92\pm1$ & $15\pm2$ & $5\pm1$\\
\hline
ERS & $H_{160} < 23$ mag & 520 & $(z_{850}-K_\mathrm{s}) - (B_{435} - z_{850}) > - 0.2
\mathrm{\ or\ } z_{850} - K_\mathrm{s} > 2.5$ & $84\pm4$ & $38\pm3$ & $6\pm3$\\
ERS & $H_{160} < 24$ mag & 1009 & $(z_{850}-K_\mathrm{s}) - (B_{435} - z_{850}) > - 0.2
\mathrm{\ or\ } z_{850} - K_\mathrm{s} > 2.5$ & $90\pm2$ & $31\pm2$ & $3\pm1$\\
ERS & $H_{160} < 25$ mag & 1768 & $(z_{850}-K_\mathrm{s}) - (B_{435} - z_{850}) > - 0.2
\mathrm{\ or\ } z_{850} - K_\mathrm{s} > 2.5$ & $86\pm1$ & $30\pm2$ & $4\pm1$\\
\hline
\end{tabular}
\end{center}
$^{\dagger}$Only galaxies matched to counterparts in the GOODS-MUSIC
catalog with measured $K_\mathrm{s}$ magnitudes are used in the
analysis of the \textit{BzK} selection efficiency.\\
$^{\star}$For \textit{YHVz} selection a target redshift interval of
$1.5 < z < 3.5$ is used for computing the selection efficiencies
and strict contamination rates, while for \textit{BzK} selection we suppose a target
redshift interval of $1.4 < z < 2.5$.\\
\end{table*}

Guided by this qualitative analysis we define a $1.5 < z < 3.5$ selection function, \textit{YHVz}, via the combination of a diagonal and a vertical limit as follows:
\begin{equation}
\begin{split}
Y_{105/098} - H_{160} > m \times (V_{606} - z_{850}) + b \\
\mathrm{\ or\ } \mathrm{\ } Y_{105/098} - H_{160} > c \mathrm{\ .}
\end{split}
\end{equation}
Maximization of the selection rate ($S$) over this parameter space with the constraint that the $z < 1$ low redshift interloper contamination rate ($C_{z < 1}$) not exceed 5\% returned the selection parameters $m=1.1$, $b=-0.25$, and $c=1.0$ for the HUDF09, and $m=1.0$, $b=-0.1$, and $c=1.5$ for the ERS.  Specifically, for the HUDF09:
\begin{equation}
Y_{105} - H_{160} > 1.1 \times (V_{606} - z_{850}) -0.25\\
\mathrm{\ or\ }
\mathrm{\ } Y_{105} - H_{160} > 1
\end{equation}

and for the ERS:
\begin{equation}
Y_{098} - H_{160} > V_{606} - z_{850} -0.1 \\
\mathrm{\ or\ }
\mathrm{\ } Y_{098} - H_{160} > 1.5 \mathrm{\ .}
\end{equation}

These optimal \textit{YHVz} selection criteria identify $96\pm1$\% and $92\pm1$\% of galaxies in our target redshift interval of $1.5 < z < 3.5$ to $H_{160} < 27$ mag in the HUDF09 and $H_{160} < 25$ mag in the ERS, respectively.  The corresponding strict contamination rates ($C$) from interlopers at lower and higher redshifts ($z < 1.5$ or $z > 3.5$) are only $15\pm2$\% in each case, with contamination from $z < 1$ low redshift interlopers ($C_{z < 1}$) of only $5\pm1$\% (as constrained during the optimization process).  The quoted uncertainties were computed according to the widths of the (median centered) intervals containing 68\% of output $S$, $C$, and $C_{z<1}$ values recovered from a series of 1000 Monte Carlo simulations (with replacement) in which the colors and redshifts of galaxies were randomly shifted according to their estimated errors---the errors in galaxy colors defined according to the Poissonian errors in the underlying broad-band fluxes, and the redshift errors set to the (scaled) values from ZEBRA+ (see Section \ref{photzs}) for galaxies with photometric redshifts only (and zero for those with spectroscopic redshifts).  This resampling process also accounts naturally for the binomial errors (Cameron 2010) in the measured selection and contamination rates.

\subsection{Trade-Offs between Sample Completeness and Sample Purity}
\label{newseccmc2}

The efficiency of \textit{YHVz}
selection is ultimately limited by intrinsic variations in the stellar population ages, star formation rates, and metallicities of galaxies, which preclude a perfect division by redshift on a single color-color diagram.  Comparison of our
optimal selection criteria against the model evolutionary tracks
presented in Fig.\ \ref{fig4} demonstrates a necessary
``trade-off'' between high completeness for both active and passive
systems at $z \ga  3$, and exclusion of active
systems at $z < 1.5$.  Photometric errors in the measured colors of galaxies contribute only modestly to the recovered selection and contamination rates due to the unprecedented depth of the HUDF09/HUDF and ERS/GOODS imaging.  Specifically, photometric errors can be excluded (above a 90\% confidence level) as the cause of \textit{YHVz} misclassifications for over 75\% of all galaxies with conflicting spectroscopic redshifts in our master source catalogs, and for over 55\% of galaxies with conflicting photometric redshifts.  We explicitly note the existence of one object in
the HUDF09, and 3 objects in the ERS, with confident \textit{YHVz} classifications
`strongly' in
conflict with their measured spectroscopic redshifts (i.e., offset by more
than 0.5 mag to the wrong side of our selection limits).  The one strongly conflicting object in the
HUDF09 (UDF0045) is a compact system at $z_{\mathrm{spec}} = 1.216$
with a spectral classification of ``broad-line AGN'' (Grazian et
al.\ 2006).  Of the 3 strongly
conflicting objects in the ERS, one has an ``AGN'' spectral class
(ERS2179, $z_{\mathrm{spec}}=1.617$), and 2 are flat spectrum sources
with noisy spectra for which we disagree with the confidence levels assigned
to the redshifts (ERS2585, $z_{\mathrm{spec}}=1.883$; ERS0463,
$z_{\mathrm{spec}}=0.117$).

The dependence of selection efficiency on apparent magnitude is an important consideration for the application of any color-selection criteria (e.g.\ for identifying specific high redshift galaxy types for spectroscopic follow up).  For sub-samples of our HUDF09 and ERS sources at brighter magnitude limits the \textit{YHVz} selection efficiencies remain stable but the contamination rates increase marginally.  For instance, at $H_{160} < 25$ mag in the HUDF09 and $H_{160} < 23$ mag in the ERS we recover contamination rates of $C = 17\pm4$\% and $21\pm$4\%, and $C_{z < 1} = 8\pm2$\%  and $10\pm3$\%, respectively.  This is an expected consequence of the higher ratio of low redshift sources to high redshift sources in the brighter sub-samples, given the intrinsic limitations of redshift determination in a single color-color diagram.  The dependence of \textit{YHVz} selection efficiency on apparent magnitude in our HUDF09 and ERS source catalogs is further quantified in Table \ref{table1}.

\begin{figure*}
\epsscale{1.15}
\plotone{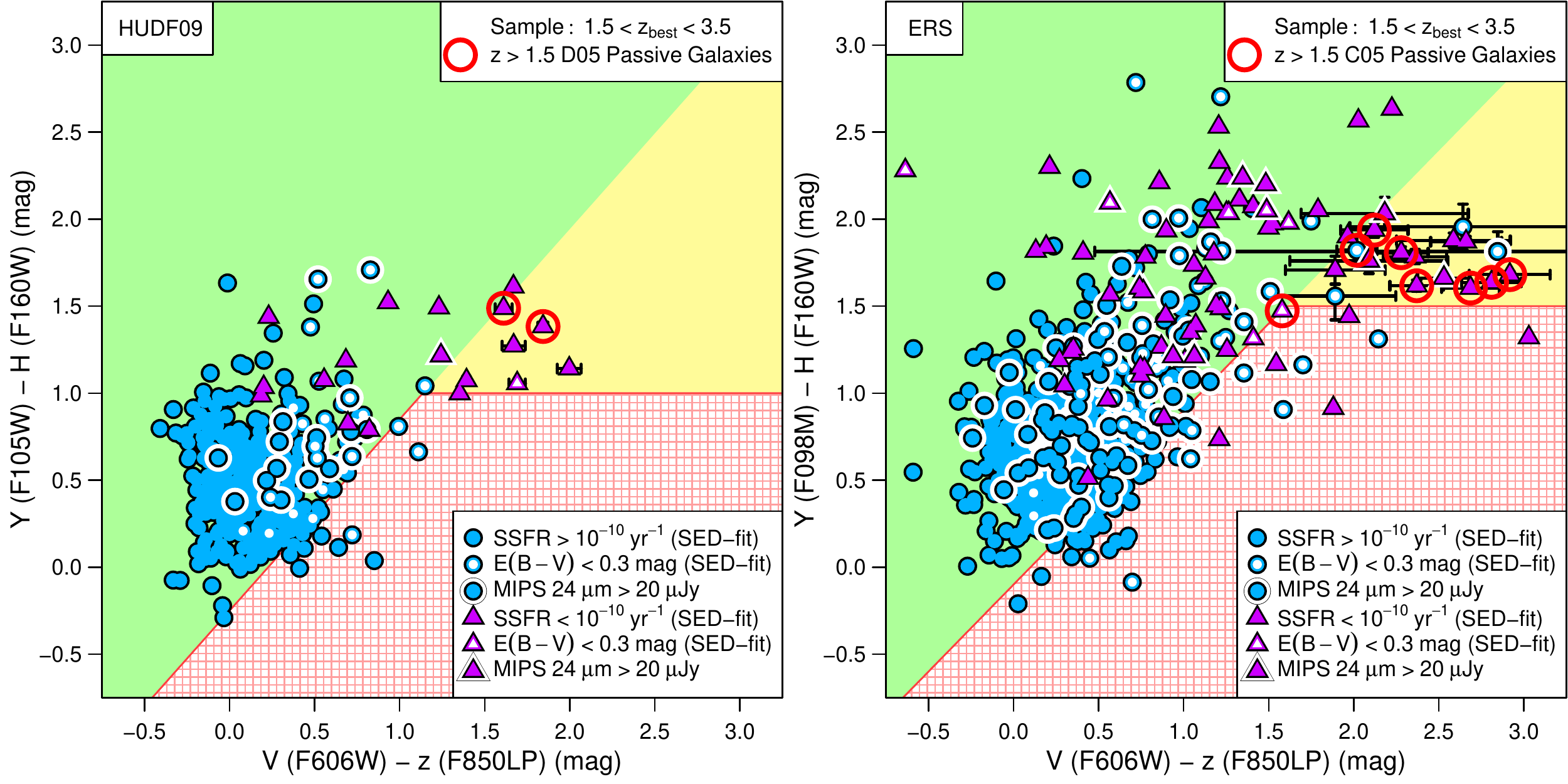}
\caption{\textit{Passive YHVz} color-selection of passively-evolving galaxies at $1.5 < z < 3.5$ in the \textit{Hubble Space
    Telescope} ACS plus WFC3 imaging of the HUDF and GOODS-South
  fields from the HUDF09 (\textit{left column}) and ERS (\textit{right
    column}) programs.  The $Y_{105/098}-H_{160}$ vs.\ $V_{606}-z_{850}$ color-color distributions of galaxies at these redshifts with SED-fit classified actively
  star-forming stellar populations are marked with blue circles and passively-evolving stellar populations with purple triangles.  Galaxies with SED-fits requiring high dust reddening values ($E(B-V) > 0.3$ mag) are marked with small white dots, and strong MIPS 24 $\mu$m sources ($F_{24\mu \mathrm{m}} > 20$ $\mu$Jy) are marked with black circles.  Yellow shading is used to highlight the sub-region of color-color space defined by our \textit{passive}
\textit{YHVz} selection criteria.  The photometric errors in the measured colors of $1.5 < z < 3.5$ galaxies within this region are illustrated by the corresponding error bars, and the positions of spectroscopically-confirmed passive galaxies at $z > 1.5$ from the \textit{pBzK} catalogs are highlighted with red circles.}
\label{fig5}
\end{figure*}

\subsection{A Comparison of Performance}
\label{newseccmc3}
\begin{figure*}
\epsscale{1.15}
\plotone{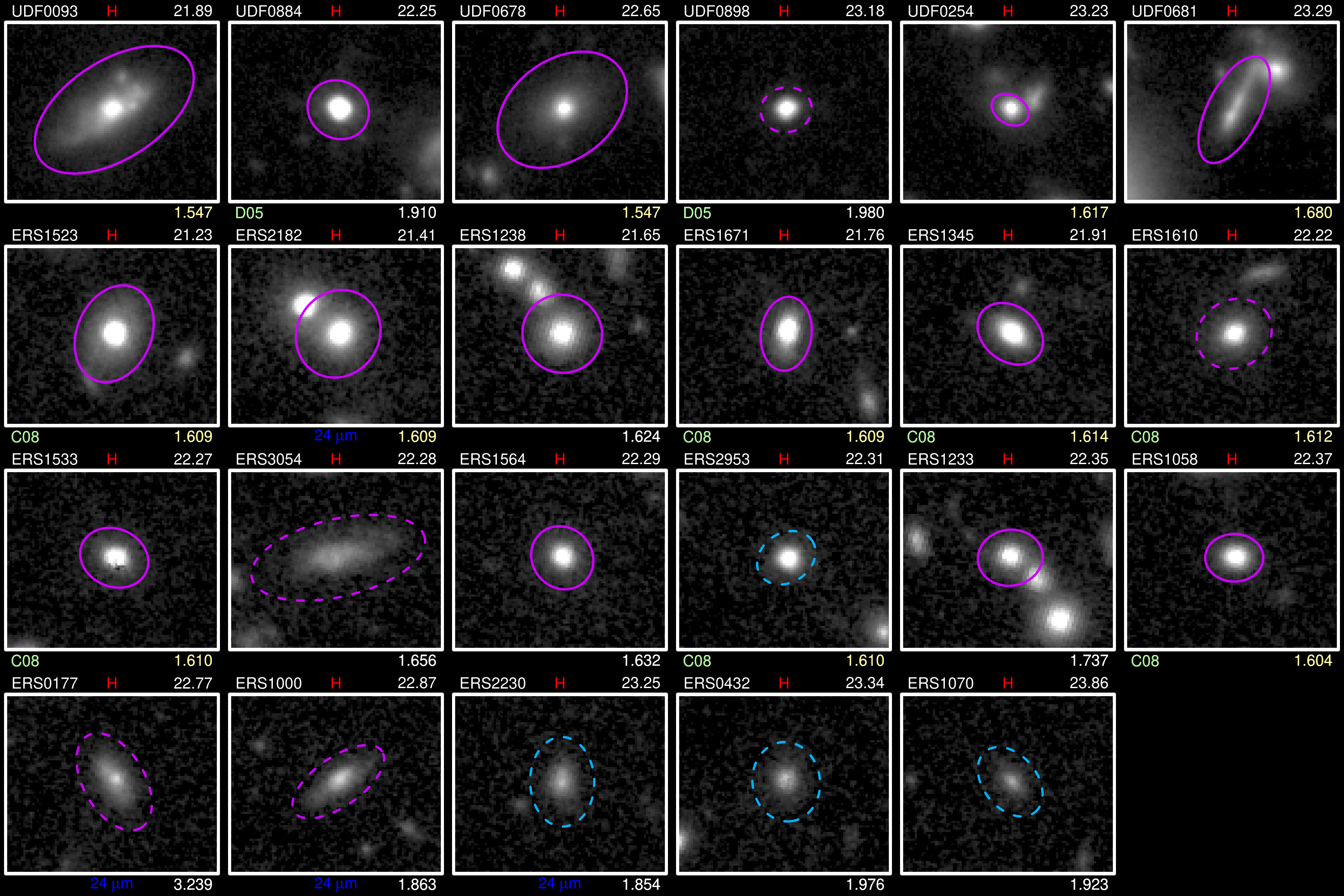}
\caption{$H_{160}$ postage stamp images of \textit{passive YHVz} galaxies at $1.5 < z < 3.5$ in the HUDF09
  (\textit{top row})
  and ERS (\textit{bottom three rows}) WFC3 imaging of the HUDF and GOODS-South
  fields.  In each case the elliptical Kron aperture identified for
  that galaxy is overlaid on the image, with the ellipse color revealing whether its best-fit SED template was indicative of a passively-evolving (purple) or star-forming (blue) stellar population.  Solid ellipses are used to denote galaxies classified within the \textit{passive YHVz} selection area at $>$90\% confidence given the photometric errors on their colors, while broken ellipses are used for lower confidence classifications.  The corresponding $H_{160}$
  apparent magnitude and best available redshift for each source are noted
  in the top and bottom right-side corners of each postage stamp image,
  respectively, with spectroscopic redshifts highlighted in yellow.  Additional notes are added for sources spectroscopically-confirmed as passively-evolving
  systems in the \textit{pBzK} Cimatti et al.\ or Daddi et al.\ samples, or alternatively, for sources identified as strong MIPS 24 $\mu$m detections ($F_{24\mu \mathrm{m}} > 20$ $\mu$Jy) in the GOODS-MUSIC catalog.}
\label{fig6}
\end{figure*}

We  have compared the efficiency of \textit{YHVz} $1.5 < z < 3.5$
selection against the \textit{BzK} $1.4 < z < 2.5$ selection of Daddi
et al.\  (2004). To this purpose, we have computed the corresponding \textit{BzK} selection and
contamination rates for all galaxies in our master source catalogs well-matched to GOODS-MUSIC objects
with measured $K_\mathrm{s}$ magnitudes.  For \textit{BzK} we obtain $S = 84\pm1$\% and $86\pm1$\%, $C = 30\pm2$\%
and $30\pm2$\%, and $C_{z < 1} = 3\pm1$\% and $4\pm1$\% to $H_{160} < 27$ mag in the
HUDF09 and $H_{160} < 25$ mag in the ERS,
respectively (as summarized in Table \ref{table1}).  Comparatively,
\textit{YHVz} selection offers a more efficient mechanism for high redshift galaxy selection with a significantly
more precise sampling of the target redshift interval than \textit{BzK} (i.e., lower
strict contamination rates).  

A further distinction
between \textit{YHVz} and \textit{BzK} is that the former fully exploits the very high angular resolution and uniformly high sensitivity of both the WFC3 and ACS, whereas the latter requires
$K_\mathrm{s}$ imaging, not available, for large galaxy samples, at a similarly high resolution.
As a result, only \textit{YHVz} selection allows the confident photometric identification of high
redshift galaxies with neighbors at angular separations below the
confusion limit of ground-based $K_\mathrm{s}$-band imaging. 
This enables more detailed studies of the structural assembly and star formation properties of galaxies  at these crucial epochs.

\subsection{Passive YHVz}\label{pYHVz}

According to the stellar population evolution models presented in Fig.\
\ref{fig5}, galaxies at $1.5 < z < 3.5$ with actively star-forming and passively-evolving stellar
populations occupy distinct regions of the $Y_{105/098}-H_{160}$ vs.\ $V_{606}-z_{850}$ color-color plane.  In particular,
high redshift passively-evolving systems are expected to exhibit
much redder $V_{606}
- z_{850}$ and $Y_{105/098} - H_{160}$ colors than star-forming systems
at the same epochs.  This is consistent with the observed colors of 
spectroscopically-confirmed passive galaxies at
$1.5 < z < 3.5$ from the Daddi et al.\ and Cimatti et al.\  samples, as shown in Fig.\ \ref{fig5}.  Interestingly, all but one of these confirmed passive galaxies can be identified simply
via an extension of our optimal \textit{YHVz} selection criteria to define a
\textit{passive YHVz} selection analagous to the \textit{pBzK}
object class.  Specifically, for the HUDF09:

\begin{equation}
Y_{105} - H_{160} < 1.1 \times (V_{606} -
z_{850}) - 0.25\\
\mathrm{\ and\ } \mathrm{\ } Y_{105} - H_{160} > 1
\end{equation}

and for the  ERS:

\begin{equation}
Y_{098} - H_{160} < V_{606} - z_{850} - 0.1 \\
\mathrm{\ and\ } \mathrm{\ } Y_{098} - H_{160} > 1.5 \mathrm{\ .}
\end{equation}

In addition to the well-studied \textit{pBzK} objects, our \textit{passive YHVz}
selection criteria identify an additional 4 and 10 galaxies with 
photometric, and/or spectroscopic, redshifts of $1.5 < z < 3.5$ in the HUDF09
and ERS, respectively.  (The one \textit{passive YHVz} galaxy  with a spectroscopic redshift not from either the considered \textit{pBzK} catalogs,  ERS2182, exhibits only a
tentative O\textsc{II} emission line detection at the upper wavelength limit of
its VLT/FORS2 optical spectrum from the GOODS program; Vanzella et al.\ 2005,
2006, 2008).  Photometric errors in the colors of these additional passive candidates enable us to robustly assign 8 (4 from the HUDF09, and 4 from the ERS)  to the \textit{passive YHVz} class above a 90\% confidence level.

According to the Calzetti (2001) model, dust reddening should move galaxies at $z \sim 2$ near-parallel to our \textit{passive YHVz} boundary, implying minimal contamination of our passive sample from dust reddened, star-forming objects at $1.5 < z < 3.5$. Indeed, the best-fit model templates to the observed SEDs of all confident \textit{passive YHVz} galaxies at $1.5 < z < 3.5$ in our catalogs were representative of very low specific star formation rates (SSFR $<$$10^{-10}$
yr$^{-1}$), with only one galaxy requiring a high dust extinction value ($E(B-V) > 0.3$ mag) for an optimum fit.  Only 3 of the 10 additional \textit{passive YHVz} candidates in the ERS not in either \textit{pBzK} sample have ZEBRA+ SEDs favouring models with intermediate SSFRs ($10^{-10}$$<$ SSFR $<$$10^{-9}$
yr$^{-1}$) and high dust reddening values; all 3 enter our selection below the 90\% confidence level.  Similarly, 4 of the 17 total (\textit{pBzK} plus new) \textit{passive YHVz} systems in the ERS were identified as strong MIPS 24 $\mu$m sources ($F_{24\mu \mathrm{m}} > 20$ $\mu$Jy) in the GOODS-MUSIC catalog (see Fig.\ \ref{fig5}). The detection of strong 24 $\mu$m emission from galaxies at these epochs provides an alternative indicator of the presence of dust-obscured star-formation (cf.\ Yan et al.\ 2004), modulo any contribution from AGN activity; again, of the four 24 $\mu$m sources contaminating our \textit{passive YHVz} catalogue, three make our selection only at low confidence levels.

\begin{figure*}
\epsscale{1.15}
\plotone{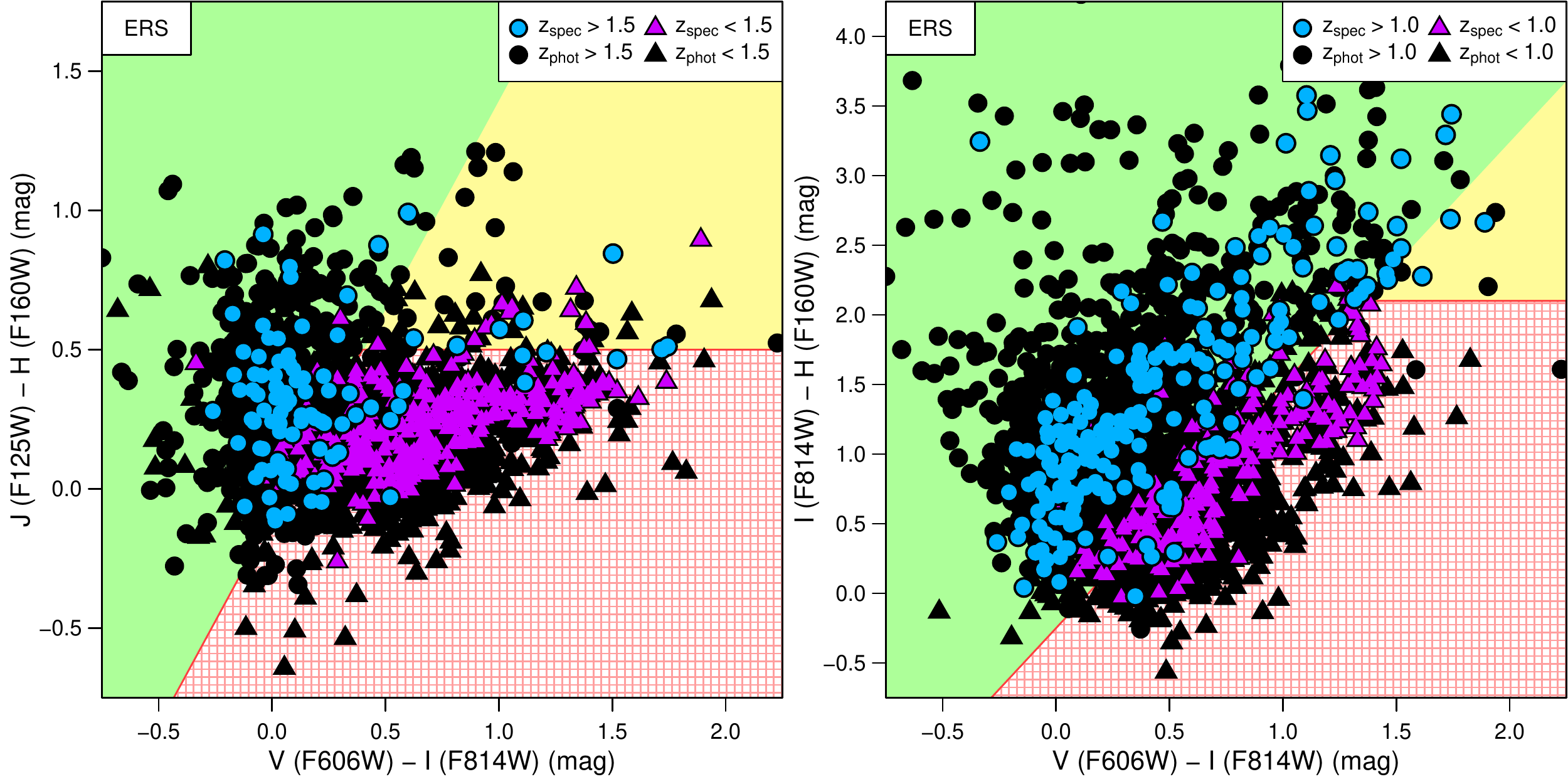}
\caption{Potential strategies for color-selection of high redshift galaxies in 
  combinations of WFC3 and ACS filters from the CANDELS/Wide program (excluding $Y_{105}$ which is available only for the GOODS-N field).  Specifically, we demonstrate
the separation of galaxies
  at $z > 1.5$ (circles) and $z < 1.5$ (triangles) in $J_{125}-H_{160}$  vs.\ $V_{606}-I_{814}$ color-color
  space (\textit{left hand panel}), and at $z > 1$ (circles) and $z < 1$ (triangles) in $I_{814}-H_{160}$ vs.\ $V_{606}-I_{814}$ color-color
  space (\textit{right hand panel}), using our ERS $H_{160} <
  25$ mag sample.  In each case, colored symboles are used if the redshift
  is spectroscopic, rather than photometric.  The optimal \textit{JHVI} and \textit{VIH} selection
  criteria we derive
  for the high redshift population in each filter combination are indicated by the
  union of the green- and yellow-shaded areas.  As described in Section \ref{mct} the $I_{814}$
  filter magnitudes of these ERS sources were estimated by
  interpolating the best-fit SED template for each galaxy against the
  $I_{814}$ transmission function.}
\label{fig7}
\end{figure*}

In Fig.\ \ref{fig6} we present WFC3 $H_{160}$ postage stamp images
of all \textit{passive YHVz} galaxies at $1.5 < z < 3.5$ from our
HUDF09 and ERS master source catalogs.  Visual inspection confirms that these are typically compact systems with highly
centrally-concentrated light profiles and smooth morphologies,
in agreement with previous studies of this galaxy class (cf.\ Van Dokkum et al.\ 2004, 2010; Daddi et
al.\ 2005; Trujillo et al.\ 2006; Cimatti et al.\ 2008; Franx et al.\ 2008; Toft et
al.\ 2009; Damjanov et al.\ 2009; Smozoru et al.\ 2010).  However, at least 6 of these
\textit{passive YHVz} galaxies exhibit morphologies suggestive of the presence of either a
stellar disk with or without clumpy sub-structure (UDF0681, ERS0177, ERS1000, UDF0093, ERS3054).  The presence of disk-like morphologies
within a subset of passive $z \sim 2$ galaxies has previously been
reported by Stockton et al.\ (2008), McGrath et al.\ (2008), and
Cassata et al.\ (2010).

We stress that the {\it passive YHVz} selection criteria (similarly to, for example, the {\it pBzK} criterion), may not return the {\it complete}  sample of passively evolving galaxies at the targeted redshifts.
At face value, there are galaxies with very low SSFRs computed with ZEBRA+ SED fits, in both the HUDF09 and ERS
samples at $1.5 < z < 3.5$, which reside outside our \textit{passive YHVz} limits, occupying a region of intermediate color-color space shared with dusty, star-forming sources.  Given that this `green valley' region is consistent with the impact of moderate-to-high dust reddening on star-forming systems at $1.5 < z < 3.5$, however, we suspect that the inherent degeneracy between the choice of reddening value and stellar population template during SED-fitting is responsible for the under-estimation of SSFRs in many of these additional passive candidates.  

\subsection{Other HST-based Criteria for Selecting $z > 1.5$ Galaxies: \\ The \textit{JHVI} Criterion}\label{mct}

Motivated by the desire to capitalize on other HST high-$z$  surveys employing similar but not identical passbands, 
we have explored the performance of other color selection criteria in isolating galaxies in the $z > 1.5$ redshift regime.
In particular, the CANDELS team is currently in the process of surveying a total area of $\sim$800 arcmin$^2$ on the sky with WFC3/IR and ACS with a two-tier observing strategy featuring a `wide, but shallow' component (CANDELS/Wide) and a `deep, but narrow' component (CANDELS/Deep; as described in Grogin et al.\ 2011 and Koekemoer et al.\ 2011).  CANDELS/Deep features observations of regions in the two
GOODS fields in each of the
$Y_{105}$, $J_{125}$, and $H_{160}$ filters, supplementing the
existing GOODS $B_{435}$, $V_{606}$, $I_{775}$, and $z_{850}$ ACS images.
Hence, our \textit{YHVz} selection can be readily applied for the study of
$1.5 < z < 3.5$ galaxies in this deep dataset.
CANDELS/Wide, however, will feature shallower imaging of regions from the GOODS-N \& S, COSMOS, EGS, and UKIDSS/UDS fields in only the $V_{606}$, $I_{814}$, $J_{125}$, and $H_{160}$ filters, precluding
application of our  \textit{YHVz} selection criteria (except in the GOODS-N region for which $Y_{105}$ data is available from the ERS and CANDELS/Deep programs).  

To estimate the efficiency of other potential color-color
selection strategies for CANDELS/Wide, we first estimate $I_{814}$
magnitudes for all objects in our ERS $H_{160} < 25$ mag source catalog by
interpolating the best-fit SED template for each galaxy against the
$I_{814}$ transmission function. The wavelength coverage of the
$I_{814}$ filter is significantly longer and redder than that of the $I_{775}$
filter employed in the GOODS ACS program, providing
deeper imaging in this part of the spectrum for a given exposure
time.  Using these interpolated $I_{814}$ magnitudes we contrast the distributions of galaxies at
$z > 1.5$ and $z < 1.5$ in our ERS
$H_{160} < 25$ mag sample in $J_{125}-H_{160}$ vs.\ $V_{606} - I_{814}$ 
color-color space in Fig.\
\ref{fig7}.  

 By repeating
the optimization procedure employed in Section \ref{colorcolor} we
recover the following \textit{JHVI} selection criteria:
\begin{equation}
J_{125} - H_{160} > 1.5 \times (V_{606} - I_{814}) -0.1 \\
\mathrm{\ or\ } \mathrm{\ } J_{125} - H_{160} > 0.5 \mathrm{\ .}
\end{equation}

It is clear from visual inspection
that the distinction between galaxies above and below $z\sim1.5$ in this \textit{JHVI} parameter
space is less efficient than that demonstrated earlier for our \textit{YHVz} selection. Nonetheless, these criteria do identify  $77\pm2$\% of ERS $H_{160} < 25$ mag galaxies at
$z > 1.5$ with a strict ($z < 1.5$) contamination rate of $26\pm1$\% and a $z
< 1$ interloper rate of only $5\pm1$\% (as constrained in the optimization).  
Thus, despite the lower performance relative to our main \textit{YHVz} selection criteria, this alternative \textit{JHVI} selection still provides a useful tool for the identification of $z>1.5$ galaxies.

\subsection{Lowering the Low-$z$ Limit: \\ \textit{VIH} Galaxies Above and Below $z=1$}

As a final remark, we note that by modifying the goals of the selection to $z >
1$ galaxies one can identify more efficient color-color criteria using
only the $V_{606}$, $I_{814}$, and $H_{160}$ filters, as shown in
Fig.\ \ref{fig7}.  The optimum \textit{VIH} criteria we recover for
the ERS source catalog are:
\begin{equation}
\begin{split}
I_{814} - H_{160} > 1.75 \times (V_{606} - I_{814}) -0.25 \\
\mathrm{\ or\ } \mathrm{\ } I_{814} - H_{160} > 2.1 \mathrm{\ .}
\end{split}
\end{equation}
These criteria detect $85\pm1$\% of ERS $H_{160} < 25$ mag galaxies at
$z > 1$ (with only $5\pm1$\% contamination from $z < 1$ interlopers).

\section{Source Catalogs}\label{sourcecatalogs}

\begin{table*}[!h]
\caption{Column Numbers and Definitions for HUDF09 and ERS Source Catalogs}
\begin{center}
\label{table}
\vskip-0.5cm
\begin{tabular}{ll}
\hline \hline
\# & Definition\\
\hline
1 & ID (in format UDF\#\#\#\# / ERS\#\#\#\#)\\
2 & Right ascension (deg) (J2000)\\
3 & Declination (deg) (J2000)\\
4 & $H_{160}$ Kron magnitude (AB)$^{\dagger}$ \\
5 & $H_{160}$ RMS error in Kron magnitude$^{\dagger}$ \\
6 & $H_{160}$ Kron radius (arcsec)$^{\dagger}$ \\
7 & $H_{160}$ Ellipticity (1-b/a)$^{\dagger}$ \\
8 & $H_{160}$ Position angle (deg) (for images aligned such that west$=$0 deg and north$=$90 deg)$^{\dagger}$ \\
9 & $H_{160}$ Stellarity parameter$^{\dagger}$ \\
10 & $B_{435}$ Kron magnitude (AB)$^{\dagger}$ \\
11 & $B_{435}$ RMS error in Kron magnitude (99 if null detection)$^{\dagger}$ \\
12 & $V_{606}$ Kron magnitude (AB)$^{\dagger}$ \\
13 & $V_{606}$ RMS error in Kron magnitude (99 if null detection)$^{\dagger}$ \\
14 & $I_{775}$ Kron magnitude (AB)$^{\dagger}$ \\
15 & $I_{775}$ RMS error in Kron magnitude (99 if null detection)$^{\dagger}$ \\
16 & $z_{850}$ Kron magnitude (AB)$^{\dagger}$ \\
17 & $z_{850}$ RMS error in Kron magnitude (99 if null detection)$^{\dagger}$ \\
18 & $Y_{105/098}$ Kron magnitude (AB) ($Y_{105}$ in HUDF09, $Y_{098}$
in ERS)$^{\dagger}$ \\
19 & $Y_{105/098}$ RMS error in Kron magnitude (99 if null detection)$^{\dagger}$ \\
20 & $J_{125}$ Kron magnitude (AB)$^{\dagger}$ \\
21 & $J_{125}$ RMS error in Kron magnitude (99 if null detection)$^{\dagger}$ \\
22 & GOODS-MUSIC match flag (0$=$unmatched, 1$=$matched)\\
23 & GOODS-MUSIC ID (0 if unmatched)\\
24 & GOODS-MUSIC area flag (0$=$not in area, 1$=$in area)\\
25 & Object appears in a high signal-to-noise region in all 7 bands flag (0$=$not in high S/N region, 1$=$in high S/N region)$^{\star}$\\
26 & Flag if GOODS-MUSIC
catalog spectroscopic redshift is strongly in conflict with the observed color (0$=$okay, 1$=$in conflict)\\
27 & Star-galaxy separation flag
(0$=$galaxy, 1$=$spectroscopically-confirmed star, 2$=$candidate star)\\
28 & ZEBRA photometric redshift\\
29 & $\chi^2$ of photometric redshift template fit\\
30 & Number of filters used in photometric redshift template fit\\
31 & \textit{Passive YHVz} selection flag (0$=$not in selection, 1$=$in selection)\\
\hline
\label{catalogs}
\end{tabular}
\end{center}
$^{\dagger}$ Parameters in columns 4-21 derived from \texttt{SExtractor} output; see Bertin \& Arnouts (1996) for detailed definitions.\\
$^{\star}$ Regions of ``high signal-to-noise'' imaging defined as the area on each frame with local median RMS less than twice the global median RMS. 
\end{table*}

The complete source catalogs employed in this study, containing all robust detections to $H_{160} < 27$ mag in the HUDF09 and $H_{160} < 25$ mag in the ERS, are publicly-available in electronic form from the following URL: \texttt{http://www.astro.phys.ethz.ch/cameronetal2011/} .  These catalogs have been carefully inspected both to remove spurious detections and to ensure selection of the optimum photometric aperture for flux measurement of each object.  A series of flags is used to indicate which objects appear within the area covered by the GOODS-MUSIC catalog, which are classified as well-matched to a counterpart in the GOODS-MUSIC catalog, which appear in a region of high signal-to-noise
imaging in all 7 ACS plus WFC3 filters, which have a GOODS-MUSIC catalog spectroscopic redshift significantly in conflict with their observed colors, which are spectroscopically-confirmed stars or unresolved stellar candidates, and which are in our \textit{passive
  YHVz} selection.  The column numbers and definitions for these catalogs are listed in Table \ref{catalogs} below.

\section{Comparison of rest-frame optical versus  rest-frame UV morphologies of $1.5 < $ \lowercase{$z$} $ < 3.5$ galaxies}\label{morph}

As a first application of our $H_{160}$-selected HUDF09 and ERS source catalogues presented in Section \ref{sourcecatalogs} we establish the most basic, and thus fundamental, facts concerning the morphologies of $1.5<z<3.5$ galaxies in the rest-frame optical window compared with the morphologies of the same galaxies in rest-frame UV light.

In recent years,  ground-based or HST/NICMOS NIR imaging sampling rest-frame optical wavelengths at $1.5 \la z \la 3.5$ has uncovered a galaxy population at these epochs rich in diversity of structure: with smooth disk/bulge-disk galaxies, and rather normal ellipticals coexisting with massive interacting/disturbed systems, as well as ultra-compact, elliptical-like galaxies, large, turbulent, clumpy disks, and compact, dispersion-dominated disks (Corbin et al.\ 2000; Daddi et al.\ 2005; Papovich et al.\ 2005; Franx et al.\ 2008; Stockton et al.\ 2008; Genzel et al.\ 2008; Conselice et al.\ 2008; Buitrago et al.\ 2008; F\"orster-Schreiber et al.\ 2009; Elmegreen et al.\ 2009). Interestingly, many galaxies at $z > 1.5$ exhibit genuinely disturbed/interacting morphologies, i.e., not only the young stellar populations visible in UV light, but, rather, most of the stellar mass in these galaxies, is arranged in an irregular manner. With the significantly higher angular resolution and uniformly high imaging sensitivity of WFC3/IR, the new HUDF09 and ERS  $H_{160}$ images offer the most detailed view to-date of the high-redshift galaxy population at rest-frame optical wavelengths.  Through visual classification of the galaxies in our $1.5 < z < 3.5$ color-selected HUDF09 and ERS samples, we thus investigate here, qualitatively,  the diversity of galaxy structure as revealed by the WFC3/IR $H_{160}$ (rest-frame optical)  images relative to the ACS $z_{850}$ (rest-frame UV) images.

The first indication of the paramount importance of the rest-frame WFC3 images in understanding the $1.5<z<3.5$ galaxy population comes from the comparison with the GOODS-MUSIC catalog discussed in Section \ref{catalog}. In most cases where no counterpart to one of our $H_{160}$ sources was found in the GOODS-MUSIC catalog, the
centroid offset that prevented automatic catalog matching was due to morphological
differences between the rest-frame UV $z_{850}$ image (used
as a basis for most GOODS-MUSIC detections) and the new rest-frame optical $H_{160}$ image provided by the WFC3.  
In a number of instances, the greater sensitivity of the WFC3
$H_{160}$ image to the smoother, underlying stellar mass
distribution provided a confident aperture definition and centroid conflicting
significantly with that adopted in the GOODS-MUSIC analysis, and no
match was possible.  It is evident that the rest-frame UV images on which the GOODS-MUSIC catalog was based  often identify, as the galaxy center, a knot of star formation  substantially displaced from the true galaxy center. This is, on the other hand, well identified in the rest-frame optical WFC3  images. 

\begin{figure*}
\epsscale{1.15}
\plotone{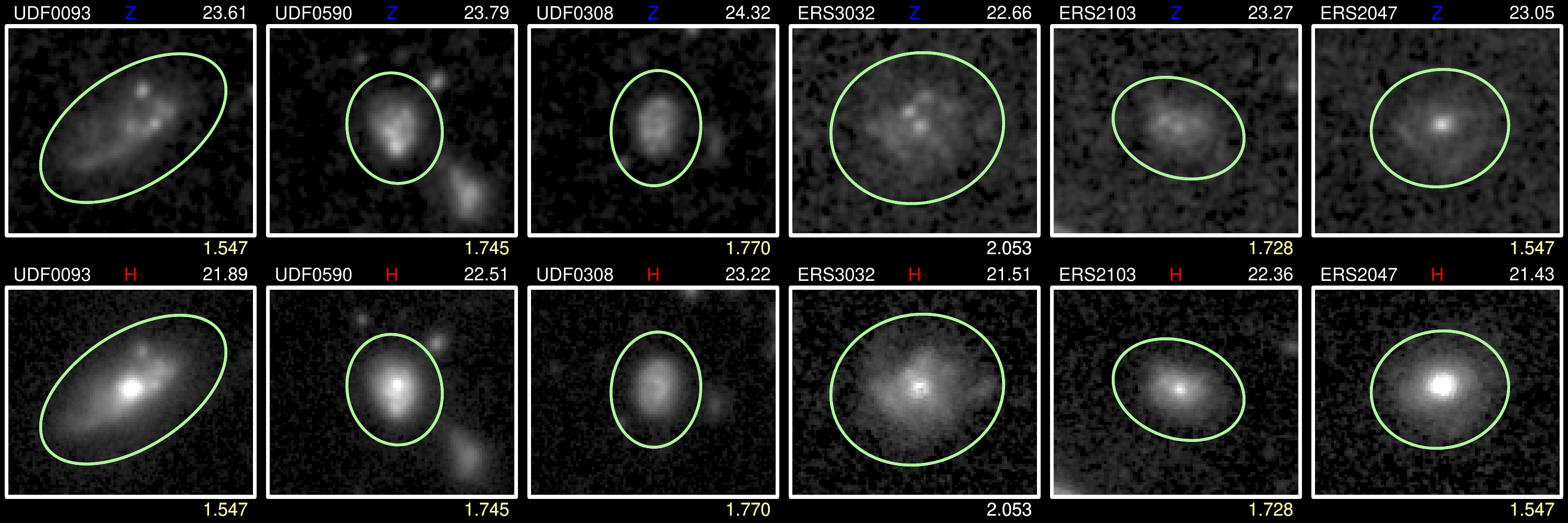}
\caption{Rest-frame UV (ACS $z_{850}$) and optical (WFC3 $H_{160}$) postage stamp images of example $1.5 < z < 3.5$ galaxies revealed as disk, or bulge/disk, systems by WFC3 (\textsc{class 1}).  In each case the elliptical Kron aperture identified for
  that galaxy is overlaid on the image.  The corresponding $z_{850}$ or $H_{160}$
  apparent magnitude and best available redshift for each source are noted
  in the top and bottom right-side corners of each postage stamp image,
  respectively, with spectroscopic redshifts highlighted in yellow.}
\label{fig8}
\end{figure*}

\begin{figure*}
\epsscale{1.15}
\plotone{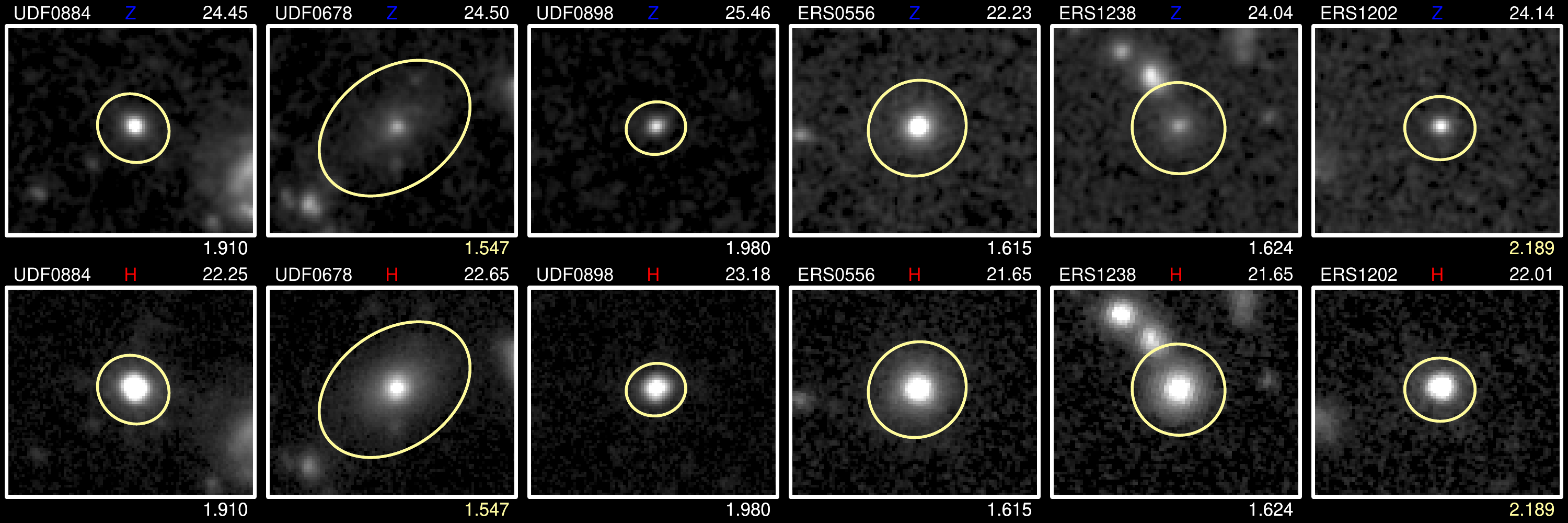}
\caption{Same as for Figure \ref{fig5}, but this time for \textsc{class 2},  $1.5 < z < 3.5$ example galaxies that show  a disk-less, spheroid-only morphology both  in the rest-frame optical WFC3 images and in the rest-frame UV ACS images. }
\label{fig9}
\end{figure*}

Also, the combination of passive-SED color-selection (\textit{passive YHVz}) and rest-frame optical morphologies (see Section \ref{pYHVz}) reveals that passive galaxies at such early epochs are indeed mostly spheroids, as shown in previous studies (e.g.\ Van Dokkum et al.\ 2004, 2010; Daddi et
al.\ 2005; Trujillo et al.\ 2006; Cimatti et al.\ 2008; Franx et al.\ 2008), but, in addition, a substantial fraction ($\sim$15-20\%) show disk morphologies.  We speculate that this population of `quenched' disks at such early epochs might be a smoking gun for disk galaxies which have been cut out of their feeding streams of cold gas, as seen in numerical simulations and expected on theoretical grounds (e.g.\ Kere\v{s} et al.\ 2005; Dekel et al.\ 2008).  Mergers of such disks would be `dry' while nonetheless involving galaxies with a dynamically-cold disk component.  Foreseen large $H_{160}$-band surveys with the HST will be instrumental in understanding what fraction of dry mergers at $z \sim 2$-3 involve disk galaxies---an important constraint for galaxy evolution models.

\begin{figure*}
\epsscale{1.15}
\plotone{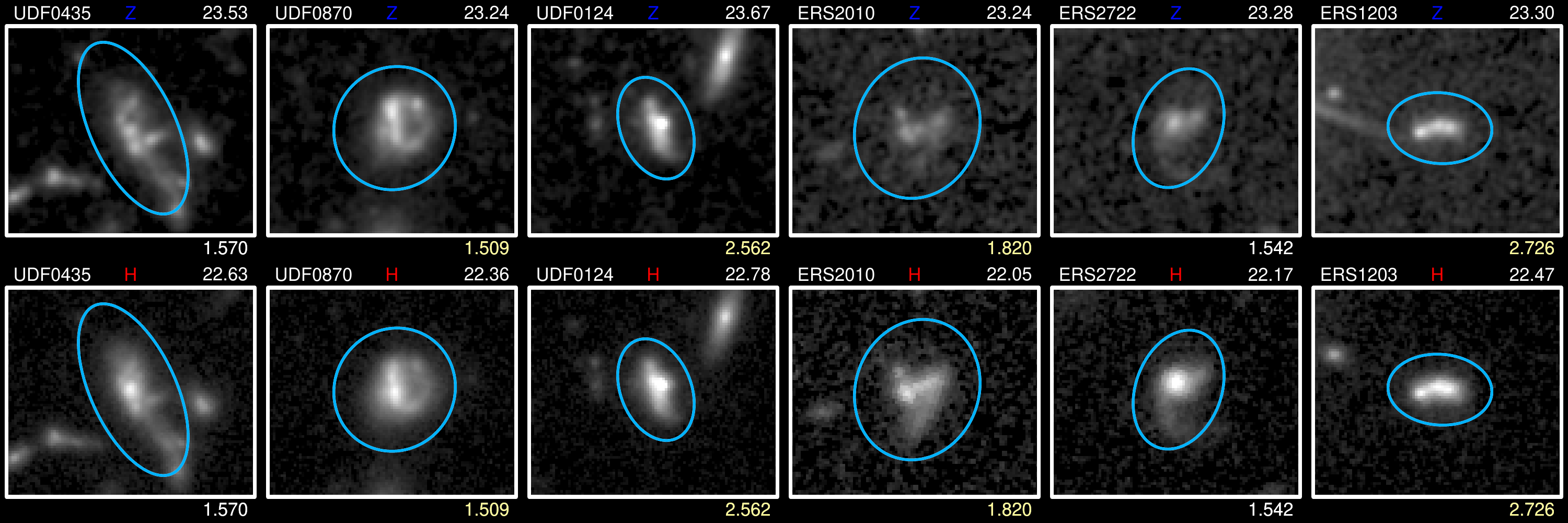}
\caption{Same as for Figure \ref{fig5}, but this time for \textsc{class 3}, $1.5 < z < 3.5$ example galaxies that keep a similar, highly irregular/disturbed morphology in rest-frame optical as in the rest-frame UV.}
\label{fig10}
\end{figure*}

\begin{figure*}
\epsscale{1.15}
\plotone{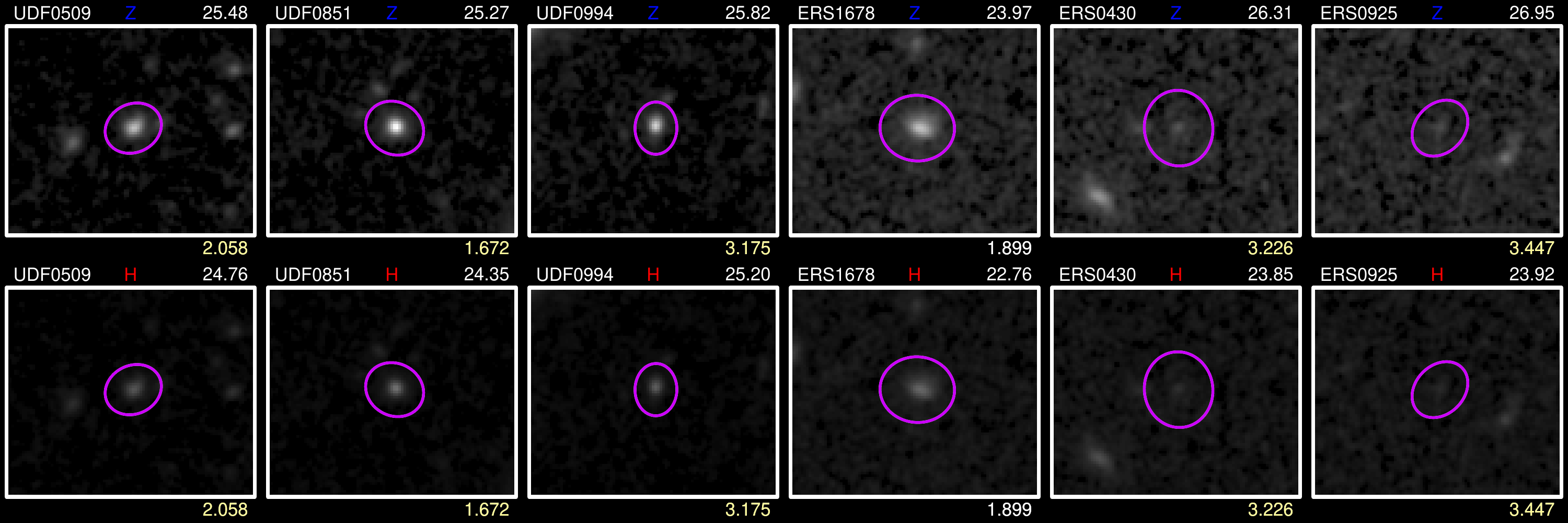}
\caption{Same as for Figure \ref{fig5}, but this time for \textsc{class 4}, $1.5 < z < 3.5$ example galaxies that have a smooth, `blob' morphology both in the WFC3 $H_{160}$ and ACS $z_{850}$ images.}
\label{fig11}
\end{figure*}

Furthermore, by comparison of the ACS and WFC3 morphologies of all galaxies in our master source catalogs within the $1.5<z<3.5$ redshift window we identify four main galaxy  sub-types. The first category (\textsc{class 1}) are galaxies which, in the $H_{160}$-band WFC3 images, reveal a regular disk morphology (with or without a bulge)---in contrast with their irregular rest-frame UV morphologies. These \textsc{class 1} galaxies are the only ones which show a morphological difference between their rest-frame UV and rest-frame optical images.  The remaining three classes do separate galaxies with very different morphologies,  but, strikingly, these  morphologies are unchanged in   rest-frame UV and rest-frame optical light. These classes are, respectively: galaxies which, both in the rest-frame UV and rest-frame optical NIR WFC3 images maintain  a regular, spheroid-like morphology (\textsc{class 2});  galaxies which retain, in the NIR WFC3 images, a  highly irregular/disturbed morphology, similar to that observed in the ACS rest-frame UV images (\textsc{class 3}); and, finally, the population of small-sized, smooth and rather symmetric `blobs' of light in both the $H_{160}$ and $z_{850}$ images for which it is unclear whether these systems are proto-disks or spheroids (\textsc{class 4}).  Examples of each class are given in Figures \ref{fig8}, \ref{fig9}, \ref{fig10}, and \ref{fig11}.

In \textsc{class 1} objects, the rest-frame optical morphology is substantially smoother, i.e., less clumpy, and also more centrally concentrated, than in the rest-frame UV.  Indeed, in such systems, the WFC3 $H_{160}$ images expose a well-defined center---in contrast with the  clumpy/asymmetric appearance of these galaxies in the rest-frame UV light. Two such galaxies in our sample (UDF0093, ERS2047) exhibit clear bulge-dominated morphologies, and a number of others (e.g.\ UDF0590, ERS3032) indicate the presence of small, central, disk light excesses, most likely small(er) bulges, consistent with counterparts identified by Papovich et al.\ (2005) and Stockton et al.\ (2008).

\textsc{Class 2} objects show the smooth and compact appearance indicated by previous studies to be typical for many spheroids at these epochs (e.g.\ Daddi et al.\ 2004, 2005; Cimatti et al.\ 2008; Franx et al.\ 2008; Damjanov et al.\ 2009).  Already Szomoru et al.\ (2010) have used the HUDF09 WFC3/IR images to establish, on firm grounds,  the compactness of one `quenched' spheroid at $z=1.98$ (UDF0884), extracted  from the Daddi et al.\ sample.  While many  galaxies in this morphological class are identified as \textit{passive YHVz} systems in our color-color selection criteria (Section \ref{pYHVz}), we stress that a non-negligible fraction of them have instead star-forming SEDs, indicating that such smooth, compact morphologies, while more typical of the  passively-evolving galaxy population at these redshifts, can also, to a smaller extent, be achieved by galaxies that are still actively forming stars at those epochs (e.g.\ ERS0556, ERS1202).
 
\textsc{Class 3} objects exhibit a wide variety of morphologies, from systems with a compact nucleus plus a highly asymmetric disk-like structure (e.g.\ UDF0124) to systems with multiple bright nuclei (e.g.\ UDF0435, ERS2010, ERS1203).  Interestingly, the disk-like structures in these systems are often warped (e.g.\ UDF0124, UDF0870), a phenomenon that is observed to occur in high-resolution, hydrodynamical  numerical simulations  in  CDM cosmology, as disk galaxies stream down the hot gas filaments of the cosmic web (Steger et al 2010, in prep.).  In a number of cases (e.g.\ UDF0435, UDF0870) there exist clear differences between the $z_{850}$ and $H_{160}$  brightnesses of specific nuclei within these galaxies,  reflecting a complex, spatially-dependent---i.e., `clumpy'---star formation activity, and thus history, in these systems.  Since these systems will most likely evolve into relatively massive galaxies by redshift zero, a substantial stellar population mixing, through mergers or internal dynamical instabilities, must occur over cosmic time to achieve the massive galaxy population with rather smooth stellar population gradients that we observe today. 

One \textsc{class 3} irregular galaxy in our sample, a lop-sided disk (UDF0870) at $z=1.509$, might exhibit a  large-scale bar. While a decline in the bar fraction with increasing redshift has been demonstrated by several studies out to redshifts of $z\sim1$ (e.g.\ Abraham et al.\ 1999; Sheth et al.\ 2008; Cameron et al.\ 2010), it is only with the newly available rest-frame optical, deep and high-resolution images of high-$z$ galaxies that a firm statement on the bar fraction at $z>1$ will become feasible. The discovery of one such system in the HUDF09 images indicates that large-scale, coherent dynamical instabilities such as bars can and do exist at such early epochs (see also the $z > 2.5$ `bar-like structures' classified in the rest-frame UV by Ravindranath et al.\ 2006), although they may be not as frequent as at redshift $z \sim 1$, given that disks at such early times appear to be, on average, substantially hotter and more turbulent than at later times (e.g.\ Genzel et al.\ 2006, 2008, 2010).

\textsc{Class 4} objects are relatively small-sized systems exhibiting smooth, symmetric morphologies, but with less centrally-concentrated light distributions than the \textsc{class 2} objects.  As such, these \textsc{class 4} objects could either be diffuse counterparts of the \textsc{class 2} spheroids, or else small proto-disks in the earliest stages of growth.  Future studies of the quantitative structural parameters and, ideally, kinematical properties of this object class are required to establish the place of these systems within the cosmic history of galaxy structural assembly.

\begin{figure}
\epsscale{1.15}
\plotone{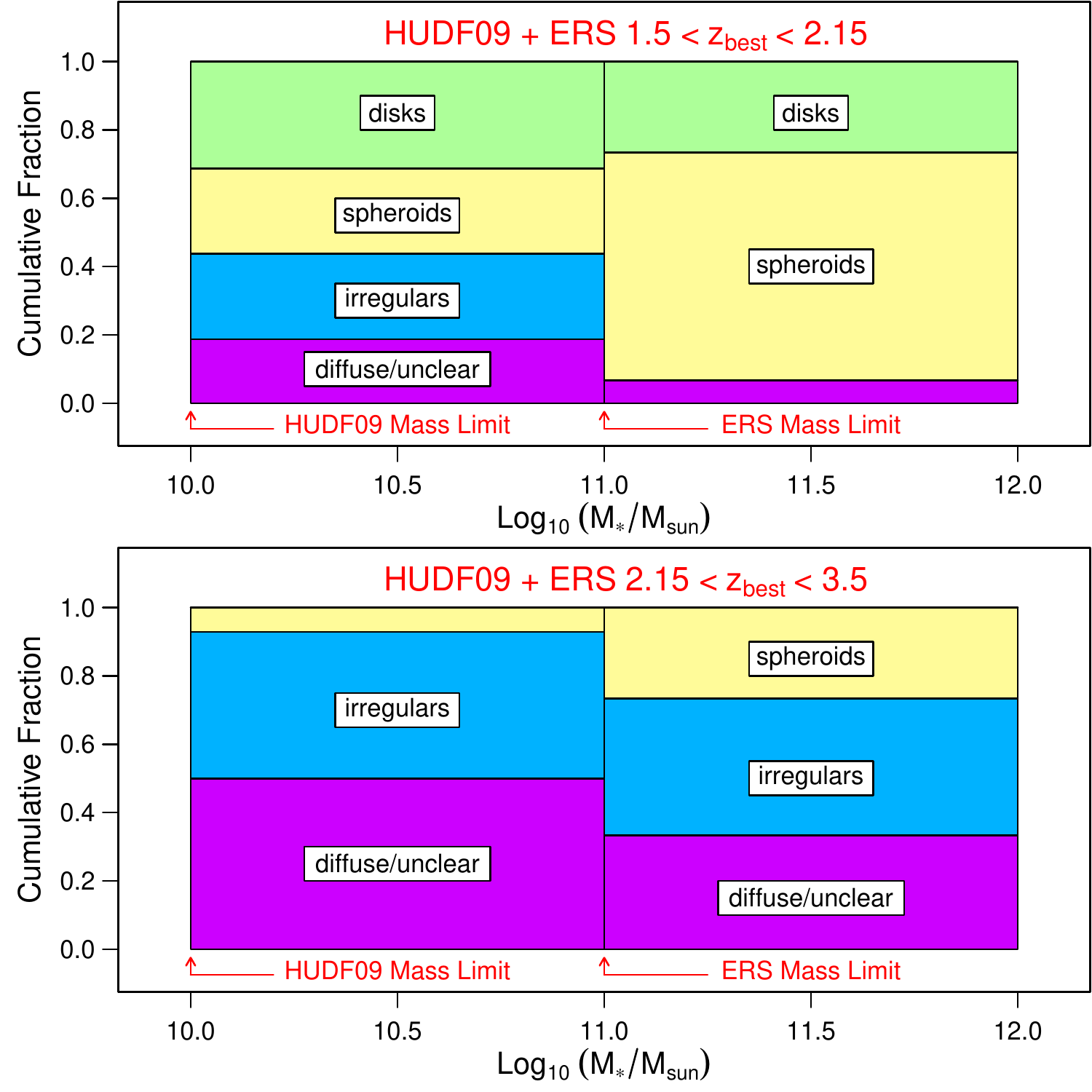}
\caption{Fractional contribution of high-$z$ morphological sub-types as a function of stellar mass at $1.5 < z < 2.15$ (\textit{top row}) and $2.15 < z < 3.5$ (\textit{bottom row}) from visual inspection of the ACS $z_{850}$ and WFC3 $H_{160}$ images of 30 galaxies at $M > 10^{10}$$M_\odot$ in the HUDF09 and 30 galaxies at $M>10^{11}$$M_\odot$ in the ERS. Green, yellow, blue, and purple histograms indicate, respectively, disks unveiled by WFC3 (\textsc{class 1}), spheroids (\textsc{class 2}), highly irregular/disturbed UV {\it and} optical morphology (\textsc{class 3}), and small-sized `blob'-like systems (\textsc{class 4}).}
\label{fig12}
\end{figure}

In Fig.\ \ref{fig12} we present the morphological mix of these galaxy types in each of the $\sim$1.2 Gyr intervals, $1.5 < z < 2.15$ and $2.15 < z < 3.5$, based upon counts in mass-limited samples of 30 galaxies at $M > 10^{10}$$M_\odot$ in the HUDF09 and 30 galaxies\footnote{Note that it is entirely by coincidence, not design, that there exist exactly 30 galaxies in each of these mass-limited samples!} at $M > 10^{11}$$M_\odot$ in the ERS (see Fig.\ \ref{fig3} for the relationship between our mass completeness limits and the distribution of galaxies in our HUDF09 and ERS source catalogs as a function of redshift).  Interestingly, as we confirm via artificial noise degradation analysis of the HUDF09 images in the Appendix to this paper, despite a $\sim$14$\times$ shorter exposure time the depth of the ERS imaging is sufficient to perform morphological classification of the (relatively crude) nature described here at comparable accuracy to that in the HUDF09.  The combined (i.e., $>$$10^{10}$$M_\odot$) fractions of objects in the four different classes are, respectively, $29\pm^{9}_{7}$\%, $45\pm^{9}_{8}$\%, $13\pm^{8}_{4}$\% and $13\pm^{8}_{4}$\% at $1.5 < z < 2.15$, and $0\pm^{6}_{0}$\%, $18\pm^{9}_{5}$\%, $41\pm^{9}_{8}$\% and $41\pm^{9}_{8}$\% at $2.15 < z < 3.5$ (with 1$\sigma$ binomial uncertainties estimated using a Bayesian approach; cf.\ Cameron 2010). Note that the evolution of these fractions likely implies substantial morphological transformations across the redshift $z \sim 2$ epoch of peak star formation activity.  We highlight, in particular, the emergence of a substantial population of regular disk morphologies, and a substantial increase in the fraction of spheroidal morphologies at epochs $z \lesssim 2$.

Importantly, the four morphological classes show different redshift and stellar mass distributions.   At $2.15 < z < 3.5$ \textsc{class 3} objects (with highly irregular morphologies also in rest-frame optical light) and \textsc{class 4} objects (with small-sized, `blob'-like morphologies) are overwhelmingly dominant at both intermediate and high mass scales, while \textsc{class 2} spheroids are largely restricted to above $10^{11}$ $M_\odot$.  At $1.5 < z < 2.15$ \textsc{class 3} and \textsc{class 4} objects are now largely restricted to intermediate masses, while the \textsc{class 2} spheroid population now dominates at high masses.  At $1.5 < z < 2.15$ the \textsc{Class 1} systems emerge and contribute $\sim$25\% of the morphological mix at both mass scales. At and around the epoch of peak cosmic star formation rate in the universe, therefore, the statistical census of the brightest galaxies indicates that the most massive of them are assembled in regular, Hubble-type elliptical and disk (plus bulge) rest-frame optical morphologies, while less massive systems often reveal, in their rest-frame optical images, clear signs of dynamical youth.

Perhaps the most striking morphological difference between these two redshift intervals is the transition across the $z \sim 2$ epoch for galaxies at the highest stellar masses, $>$$10^{11} M_\odot$, from a diversity of structural types to a dominant population of spheroids.  Although SED-fit based SSFRs suffer from large uncertainties, and our samples of galaxies with masses above $10^{11} M_\odot$ at these redshifts are only small (16 at $2.25 < z < 3.5$ and 15 at $1.5 < z < 2.25$), we note that the assembly of the massive spheroids appears to be accompanied by a rapid `quenching' of star formation.  Adopting a redshift-dependent division between star-forming and passively-evolving systems based on the SSFR necessary for a galaxy to double its stellar mass between its observed redshift and the present day, and employing our SED-fit SSFRs computed with ZEBRA+ (see Section \ref{ssfrs}), we find the fraction of star-forming galaxies at $>$$10^{11} M_\odot$ falls from $63\pm^{10}_{13}$\% at $2.25 < z < 3.5$ to $5\pm^{13}_{2}$ at $1.5 < z < 2.25$.  Using our own color-based criteria for identification of a ``passively-evolving'' SED, i.e., \textit{passive YHVz}, the fraction of quenched galaxies above $10^{10}$ $M_\odot$ increases from $6\pm^{12}_{2}$\% at $2.25 < z < 3.5$ to $60\pm^{11}_{13}$\% at $1.5 < z < 2.25$.  This is consistent with the observed decline in the star-forming fraction estimated by Feulner et al.\ (2005) for a selection of galaxies in the FORS-Deep and GOODS-South fields.  If confirmed by larger samples, these results would indicate that $z\sim2$ is the epoch when both morphological transformations first produce a significant population of massive spheroids and physical processes quench their star formation activity and move them to the red sequence, as expected in recent models (Peng et al.\ 2010).

\section{Summary}
\label{Conclusions}
The wealth of WFC3 imaging of the high redshift universe achieved or planned, coupled with archival ACS optical imaging,
is self-sufficient, without additional ground-based photometry, to robustly identify galaxies in the $1.5<z<3.5$ redshift window.
In this paper we have presented, to this aim, new, robust and efficient  color-selection criteria established primarily on the  $Y_{105/098}-H_{160}$ vs.\
$V_{606}-z_{850}$ color-color diagram (and similar ones, slightly less efficient but still valuable, which use combinations of the HST $J_{125}$, $H_{160}$, $V_{606}$,  and $I_{814}$ filters,  or simply the $V_{606}$, $I_{814}$,  and $H_{160}$ filters).  We have also shown that an additional color selection on the $Y_{105/098}-H_{160}$ vs.\
$V_{606}-z_{850}$ color-color diagram further enables the disentanglement of passively-evolving, `quenched' galaxies from actively star-forming galaxies at these redshifts. 

As a first use of the new galaxy catalog that we have produced, starting from the $H_{160}$ WFC3 images, we have performed a qualitative comparison of $1.5<z<3.5$ galaxy morphologies between rest-frame UV and rest-frame optical light.  Regular disk galaxies---with or without a bulge---are present at these epochs---the disk components being unseen or irregular in rest-frame UV images, and being clearly revealed in the rest-frame optical images.  Together with the population of smooth spheroids, these regular disk galaxies have substantial stellar masses, $>$$10^{10} M_\odot$ (or, for the spheroids, even above this threshold).  At masses $>$$10^{11} M_\odot$, the WFC3 NIR images reveal a transition of morphologies from a variety of structures at $z \gtrsim 2$ to a marked dominance of elliptical-like stellar distributions at $z \lesssim 2$.  Across the $z\sim2$ boundary the fraction of star-forming galaxies at $>$$10^{11} M_\odot$ decreases from $\sim$60\% to $\sim$5\%.  Consistent with analysis of the $z \le 1$ population (Peng et al.\ 2010), and observed here in rest-frame optical morphologies and colors, $z \sim 2$ appears therefore to be the epoch at which assembly and quenching of the massive spheroid population begins in earnest.  The WFC3 NIR images also reveal, however, that, less massive galaxies at the same epochs often keep a highly irregular morphology, also in the rest-frame optical light, as they had in the rest-frame UV light. This indicates that their previously reported irregular morphologies in rest-frame UV are not driven by  detection of star-forming populations on top of settled, underlying older stellar populations, but rather an indication of a genuine juvenile dynamical state for much of the $10^{10} < M < 10^{11}$$M_\odot$ galaxy population at these redshifts.\\

\section*{Acknowledgments}
Based on observations made with the NASA/ESA \textit{Hubble Space Telescope},
obtained from the Data Archive at the Space Telescope Science
Institute, which is operated by the Association of Universities for
Research in Astronomy, Inc., under NASA contract NAS 5-26555.  P.O.\
acknowledges support from the Swiss National Foundation (SNF).

\appendix
In this brief appendix we confirm through statistical image degradation that the limited depth of the ERS imaging with WFC3/IR is indeed sufficient to allow for confident morphological classification of $z\sim2$ galaxies in the rest-frame optical, at least to the relatively crude level of detail employed in the analysis of Section \ref{morph}.  Adopting the ultra-deep HUDF09 $H_{160}$ images of our $1.5 < z < 3.5$ master sample as the benchmark for morphological classification at these redshifts we thus aim to demonstrate a consistency of visual appearance following artificial degradation to the noise level of the ERS.  In particular, the net exposure time of the master UDF $H_{160}$ image is a factor of $\sim$14 times greater than that of the ERS $H_{160}$ image; indeed such an extraordinary depth at these wavelengths was crucial to the principal aim of the HUDF09 program---namely, the search for $z\sim8$-10 candidates appearing as $Y_{105}$/$J_{125}$ drop-outs (Bouwens et al.\ 2010a, 2011).  Consequently, the contributions of both background sky noise and shot noise in the ERS are $\sim$$\sqrt{14}$ times higher than in the HUDF09.  The final image noise in our master frames is, moreover, weakly correlated between adjacent pixels as a result of the reduction process.  Hence, rather than simply adding an independent noise component to each pixel to simulate the degradation in sky noise we attempt to at least partially approximate the impact of pixel-to-pixel correlation by degrading our UDF images instead through the stacking of blank regions of real background sky.  For the addition of shot noise, however, we simply suppose that all flux detected above the 3$\sigma$ sky noise level in the original UDF images corresponds to real object flux, and we directly resample this flux (with Poisson uncertainty) in the $\textsc{R}$ statistical package at the ERS rate.

\begin{figure*}
\epsscale{1.15}
\plotone{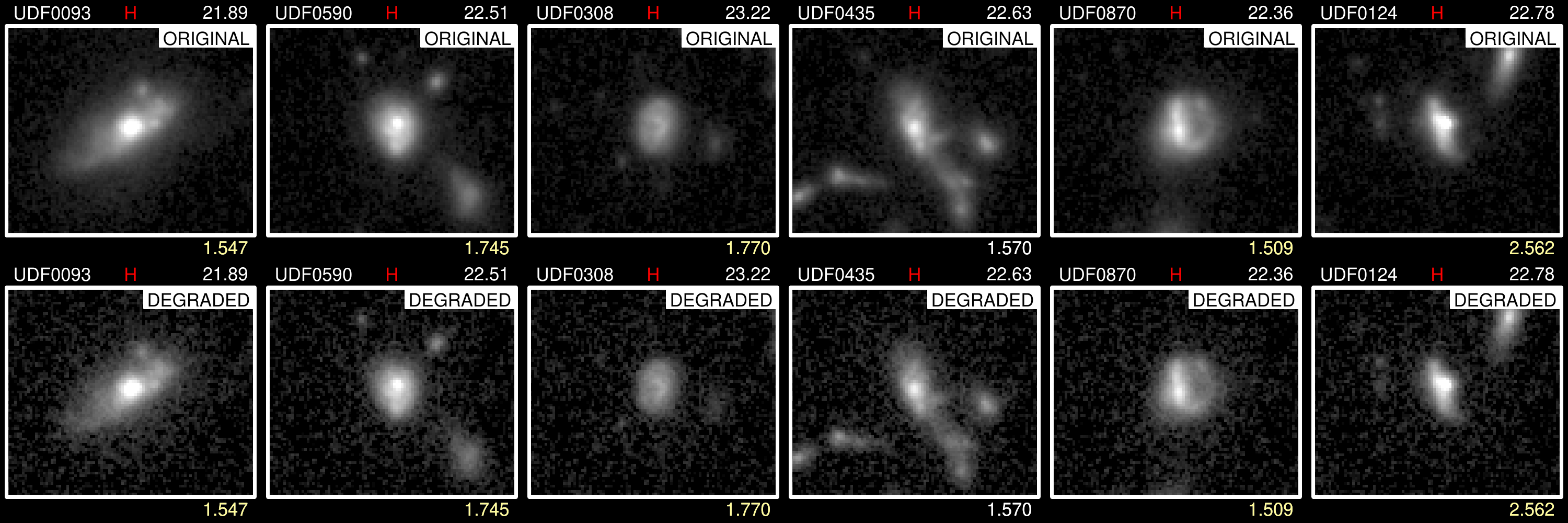}
\caption{Confirmation of the robustness of our visual morphological classifications of $1.5 < z < 3.5$ galaxies in the ERS rest-frame optical ($H_{160}$) imaging of the GOODS-South field with WFC3 despite the greater contribution from background sky and shot noise relative to the deeper HUDF09 data.  Specifically, the original ($H_{160}$) HUDF09 postage stamp images of three archetypal \textsc{class 1} (disk, or bulge/disk) systems from Fig.\ \ref{fig8}, and three archetypal \textsc{class 3} (irregular) systems from Fig.\ \ref{fig10}, are presented in the top row of this Figure for comparison against versions of these images degraded to the noise level of the ERS in the bottom row.  Although
 the faint morphological features of these galaxies appear less distinct after degradation they are still clearly recognisable as members of their assigned classes.  Hence, we argue that the depth of the ERS imaging is indeed sufficient for confident morphological classification at the (crude) level of detail we employ in the analysis of Section \ref{morph}.}
\label{fig13}
\end{figure*}

In Figure \ref{fig13} we compare the original ($H_{160}$) HUDF09 postage stamp images of three archetypal \textsc{class 1} (disk, or bulge/disk) systems from Fig.\ \ref{fig8}, and three archetypal \textsc{class 3} (irregular) systems from Fig.\ \ref{fig10}, against our new versions of these images degraded to the noise level of the ERS.  It is clear from inspection of this Figure that the impact of the additional noise is to reduce the relative prominence of certain faint and/or fine-structure features in these $z\sim2$ galaxies, but that the broad morphological properties of each system (e.g., the presence/absence of multiple nuclei, a central bulge, an outer disk, or a warped disk) remain readily discernable, and thus each galaxy remains clearly recognisable as a member of its assigned class.  Of course, studies attempting to achieve finer morphological classifications than those conducted here, or else attempting to perform quantitative `clump-science' (e.g., F\"orster-Schreiber et al.\ 2011), may well prove more sensitive to the imaging depth than our relatively crude analysis of Section \ref{morph}.  Further, one should not attempt to infer from the results of this Appendix that certain quantitative morphological analysis procedures (e.g., bulge-disk decomposition) will also prove insensitive to the difference in the HUDF09 and ERS imaging depths---such a statement clearly requires a more detailed investigation tailored to each morphological analysis tool under study.

\begin{figure*}
\epsscale{1.15}
\plotone{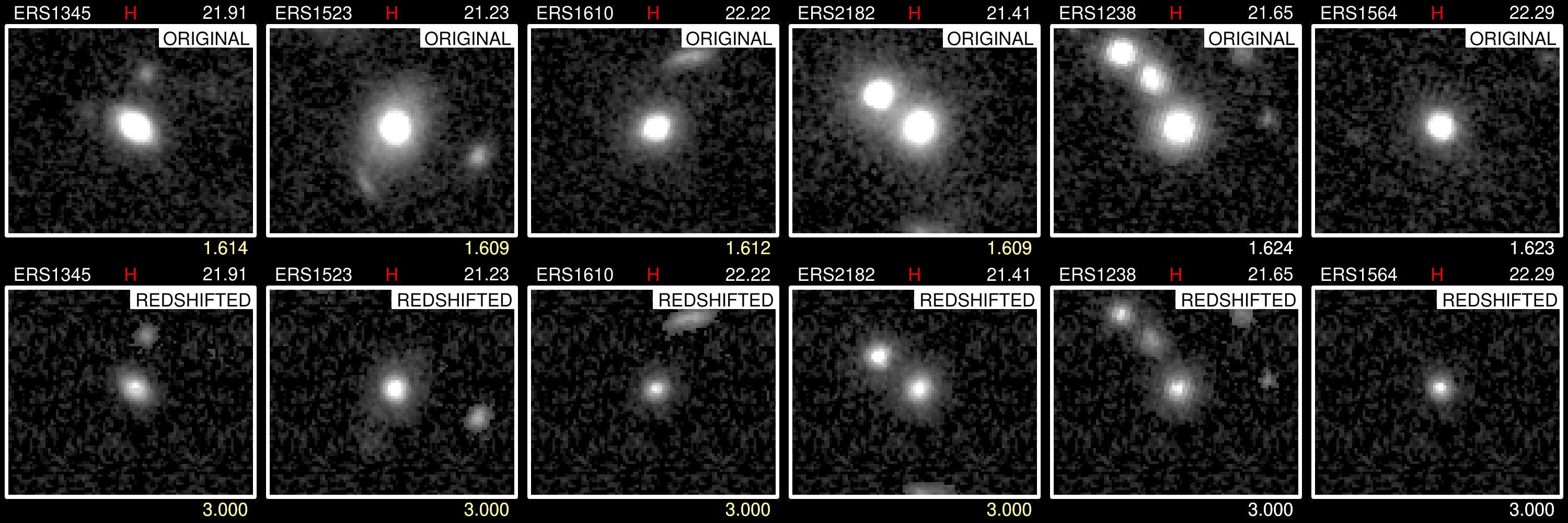}
\caption{Confirmation of the consistency of visual morphological classifications of ellipticals at $z\sim1.5$ and $z\sim3$ in the ERS rest-frame optical ($H_{160}$) imaging of the GOODS-South field with WFC3.  Specifically, the original ($H_{160}$) ERS postage stamp images of six archetypal \textsc{class 2} spheroids at $z\sim1.5$ are presented in the top row of this Figure for comparison against versions of these images artificially redshifted to $z\sim3$ in the bottom row.  The extremely centrally-concentrated light profiles of these giant ellipticals ensure that even under substantial cosmological surface brightness dimming they remain morphologically recognisable and quite distinct from the remaining three classes.  Note that in the cases of ERS2182 and ERS1238 we artificially redshift also their nearby companions, which our photometric (and/or spectroscopic) redshifts indicate to be at very nearly identical epochs.}
\label{fig14}
\end{figure*}

As a further robustness test of the morphological evolution described in Section \ref{morph} we have investigated the possibility that the decreasing fraction of massive spheroids with increasing redshift (past $z\sim2$) is an artifact of the impact of cosmological surface brightness dimming (and band-pass shifting) on our observations.  In particular, we have verified that our visual classifications of all massive elliptical galaxies in our $1.5 < z < 2.15$ sub-sample remain consistent after artificial redshifting of their $H_{160}$ images to $z=3$.  The results of this redshifting for six archetypal $z\sim1.5$ ellipticals from our sample are displayed in Fig.\ \ref{fig14}.  The extremely centrally-concentrated light profiles of these systems ensure that even under substantial cosmological surface brightness dimming they remain quite distinct in appearance from the remaining three classes.

\end{document}